\documentclass[preprint,10pt,authoryear]{elsarticle}
\usepackage{amssymb}
\usepackage{amsmath}
\usepackage{physics}

\begin{document}

\begin{frontmatter}

 \title{Between equilibrium and fluctuation: Einstein's heuristic argument and Boltzmann's principle}
 \author{Enric Pérez \corref{cor1}}
 \ead{enperez@ub.edu}
  \author{Antonio Gil}
  \ead{antoniogilmoreno2001@gmail.com}
 
 \cortext[cor1]{Corresponding author}
 
 \affiliation{organization={Universitat de Barcelona, Departament de Física de la Matèria Condensada},
            addressline={Martí i Franquès 1-11}, 
            city={Barcelona},
            postcode={08028}, 
            state={},
            country={Spain}}

\begin{abstract}
We critically revisit Einstein’s 1905 heuristic argument for lightquanta, considering its internal coherence and the scope of its applicability. We argue that Einstein’s reasoning, often celebrated for its originality,  is ambiguous because it can be understood as a fluctuation or as a comparison between equilibrium states. A historical and conceptual analysis of Einstein’s use of Boltzmann’s principle in those years reveals his evolving stance on its meaning and the role of probability, as well as his persistent doubts about the nature of radiation. We use our analysis to examine the limitations of extending the notion of Einstein's lightquanta across the electromagnetic spectrum: the relevant parameter is not the frequency, but the occupancy number.
\end{abstract}

\end{frontmatter}

\tableofcontents


\section{Introduction}
\label{sec1}

The birth of quantum mechanics has been extensively studied. One of the most significant paths to the new theory in 1925—often referred to as the ``statistical route''—can be traced back to Ludwig Boltzmann’s work and its subsequent development by Planck \citep{kuhn,darrigol,desalvo,duncanjanssen2019,navarro}.\footnote{See also the \textit{Editorial Note} in \cite{stachel1989}, pp. 134–148.} This route coincides with the formalization of statistical physics, as seen, for instance, in Albert Einstein’s early papers \citep{einstein1902,einstein1903,einstein1904} and Josiah W. Gibbs’s foundational book \citep{gibbs}. The incorporation of statistical methods into physics prompted intense debate and even misunderstandings, but ultimately led to the development of quantum statistics between 1924 and 1926 \citep{darrigol, monaldi,pereziban}. Prior to this, fundamental statistical concepts were under discussion—such as the meaning of \textit{complexion}, the probability of a microstate, and the controversial inclusion of the factor $N!$ in the treatment of ideal gases.

From a historiographical perspective, it is noteworthy how the limited mastery and trust in statistical reasoning contributed to the initial rejection of ideas that would later become central to modern physics. Such hesitation is not unusual in the history of physics; several developments that are now seen as pivotal were met with skepticism at first. Notable examples include Max Planck’s introduction of energy quanta in 1900 and Niels Bohr’s atomic model in 1913. In this paper, we focus on Einstein’s 1905 argument for light quanta. Needless to say, this argument has been extensively examined \citep{dorling, kuhn, irons, provostbracco, norton, rynorenn, duncanjanssen2019}. It is well known that Einstein regarded this as the only truly revolutionary paper among those he published in 1905 (\cite{norton}, p. 72), and several historians have emphasized the groundbreaking nature of this contribution, especially in comparison to Planck’s work \citep{kuhn,taschetto}. Nevertheless, we believe it is still worthwhile to revisit it. Virtually all prior analyses emphasize its originality, as well as its lack of rigorous justification—or even its apparent circularity. There remains no consensus on whether the argument is ultimately valid, despite the profound consequences it inspired.

It is important to recall that Einstein himself referred to the argument as ``heuristic''—- that is, ``serving as an aid to learning, discovery, or problem --solving by experimental and especially trial-and-error methods,'' according to Merriam-Webster.\footnote{Check online on January 26th.} Einstein echoed this modest view of his 1905 work more than four decades later in a widely quoted letter to his friend Michele Besso:\footnote{Einstein to Besso, December 1951. In \citep{speziali}, p. 453. Our translation.}

\begin{quote}
Fifty years of deliberate reflection have not brought me any closer to answering the question, ‘What are light quanta?’ Of course, nowadays every scoundrel believes he knows the answer — but he is merely deluding himself.
\end{quote}

To this day, the concept of photon continues to be debated. It has even argued that ``there is no such thing as a photon'' (\cite{lamb}, p. 77). Klaus Hentschel has delineated up to 12 semantic layers that overlay this idea \citep{hentschel}. There is some agreement that it makes no sense to use a naive, corpuscular view that implies localized particles traveling in space. However, numerous experiments conducted since the mid-20th century have demonstrated the corpuscular behavior of light \citep{muthukrishan}. More recently, non-locality and entanglement have once again placed photons at the center of debates on duality \citep{shalm,rieckers,villasboas}. Although it is generally accepted that a photon is an excitation of a field that is only localized when detected by a photodetector (\cite{muthukrishan}, p. 18), even today there are proponents of semiclassical views of light and limit quantization to matter \citep{rashkovskiy}. In short, the photon is far from being a concept with a universally agreed-upon definition.

We would like to contribute to this debate with our historical study. Hentschel \citep{hentschel} has already masterfully undertaken this task in what we might call a broad overview. We will focus on the statistical route that Einstein took to arrive at his conclusion. Our aim is to draw on these reflections concerning Einstein’s argument—and some of the contemporary responses it provoked—to illustrate how the history of physics offers valuable tools for engaging with physical theory and its foundations. The historical pathways that led to the formulation of a given concept may have left no visible trace—and they need not be part of the background of a physics student focused solely on mastering the discipline as it is understood today. However, analyses based on now-outdated concepts—often the result of limited tools or frameworks before the advent of new theories—can still offer perspectives, insights, and interpretive features that may remain obscured in modern physics. That is what we hope to highlight here. To what extent is it meaningful to consider a fluctuation such as the one Einstein proposed? Can Boltzmann’s principle be applied in such a system? Does it make sense to attribute entropy to a fluctuation? As we will see, Einstein himself expressed doubts regarding these very questions. Still, regarding the modern notion of photon and its applicability across the entire electromagnetic spectrum: Is it meaningful to refer to photons with wavelenghts in the radio range? What does contemporary theory have to say on the matter?

We begin this paper by outlining Einstein’s original argument (Sec. \ref{einsteinar}), followed by a review of several historical analyses it inspired (Sec. \ref{analy}). We then examine the development of Einstein’s ideas over time, focusing on the period up to shortly after the first Solvay Conference, in the autumn of 1911 (Sec. \ref{boltzmannp}). Finally, we explore (Sec. \ref{inred}) the range of applicability of Einstein’s hypothesis—first from a historical standpoint, and then in light of modern theoretical perspectives.

\section{Einstein's argument} \label{einsteinar}

Let us briefly review the sections of Einstein’s 1905 paper on light quanta that are most relevant for our discussion \citep{einstein1905}. In Section \textsection 3, Einstein examines the entropy of radiation, which can be expressed as:

\begin{equation} \label{entrsum}
S = v \int_{0}^{\infty} \varphi (\rho, \, \nu) \, d\nu.
\end{equation}
Here, $\varphi$ is a function of $\rho$ and $\nu$, that is, of the energy density (energy per unit volume and frequency) and its frequency, while $v$ denotes the volume. This decomposition of the total entropy into monochromatic components relies on the possibility of separating radiation of different colors “without expenditure of work or supply of heat” (\cite{beck2}, p. 91).\footnote{When referring to a specific page of Einstein’s papers, we cite the English edition of the \textit{Collected Papers of Albert Einstein}.} Einstein observes that the dependence of $\varphi$ can, in principle, be reduced to a single variable by noting that the entropy of radiation enclosed between reflecting walls remains unchanged under adiabatic compression. However, he chooses not to pursue this route, as his focus is on determining $\varphi$ starting from $\rho$. In the case of blackbody radiation, $\rho$ is a function of $\nu$ such that it yields the maximum entropy

\[ \delta \int_{0}^{\infty} \varphi (\rho, \nu) \, d \nu = 0 \] 
for a given total energy

\[ \delta \int_{0}^{\infty} \rho \, d \nu = 0. \]
As a consequence, 

\begin{equation} \label{lagr}
\int_{0}^{\infty} \left( \frac{\partial \varphi}{\partial \rho} - \lambda \right) \delta \rho \, \delta \nu = 0,
\end{equation}
where $\lambda$ does not depend on $\nu$. Therefore, for black-body radiation $\partial \varphi/\partial \rho$ depends on $\nu$ neither. Going back to \eqref{entrsum}, when we vary temperature (with $v=1$):

$$dS=\int_{\nu=0}^{\nu=\infty} \frac{\partial \varphi}{\partial \rho} d\rho d \nu,$$
and due to the independence of $\partial \varphi/\partial \rho$ of $\nu$:

  \[ dS = \frac{\partial \varphi}{\partial \rho}  dE. \] 
As $dE$ is the heat transfer and the process is reversible, it can be established that:
\begin{equation} \label{thermo}
dS = \frac{1}{T} dE,
\end{equation}
being $T$ the temperature. Hence: 

\begin{equation} \label{term}
\frac{\partial \varphi}{\partial \rho} = \frac{1}{T}, 
\end{equation}
that connects $\varphi$ and $\rho$. When integrating, we will use that $\varphi=0$ for $\rho=0$. This expression is what will allow us to obtain the monochromatic entropy from the energy density, which is what is measured experimentally.

In the next section, Einstein argues that the black body radiation formula:

\[\rho = \alpha \nu^3 e^{- \beta \nu / T} \]
($\alpha$ and $\beta$ are constants) is valid only for large values of $\nu/T$ (low densities). Therefore, the results obtained using it will be valid only in that region. We can write:

\[ \frac{1}{T} = - \frac{1}{\beta \nu}  \ln{ \frac{\rho}{\alpha \nu^3} }, \] 
and using relation \eqref{term} we obtain the corresponding monochromatic entropy: 

\begin{equation} \label{ewien}
\varphi (\rho, \nu) = - \frac{\rho}{\beta \nu} \left( \ln{ \frac{\rho}{\alpha \nu^3} } - 1 \right).
\end{equation}
For obtaining the value of entropy of radiation of energy $E$ with frequency between $\nu$ and $\nu + d \nu$ with volume $v$, we write ($E=\rho v d\nu$):

$$S=v \varphi(\rho,\nu)d\nu = - \frac{E}{\beta \nu} \left\{\ln \frac{E}{v \alpha \nu^{3} d \nu} - 1\right\}.$$

\begin{figure} [ht]
	\centering
		\includegraphics[trim = 5mm 20mm 10mm 20mm, clip,width=7cm]{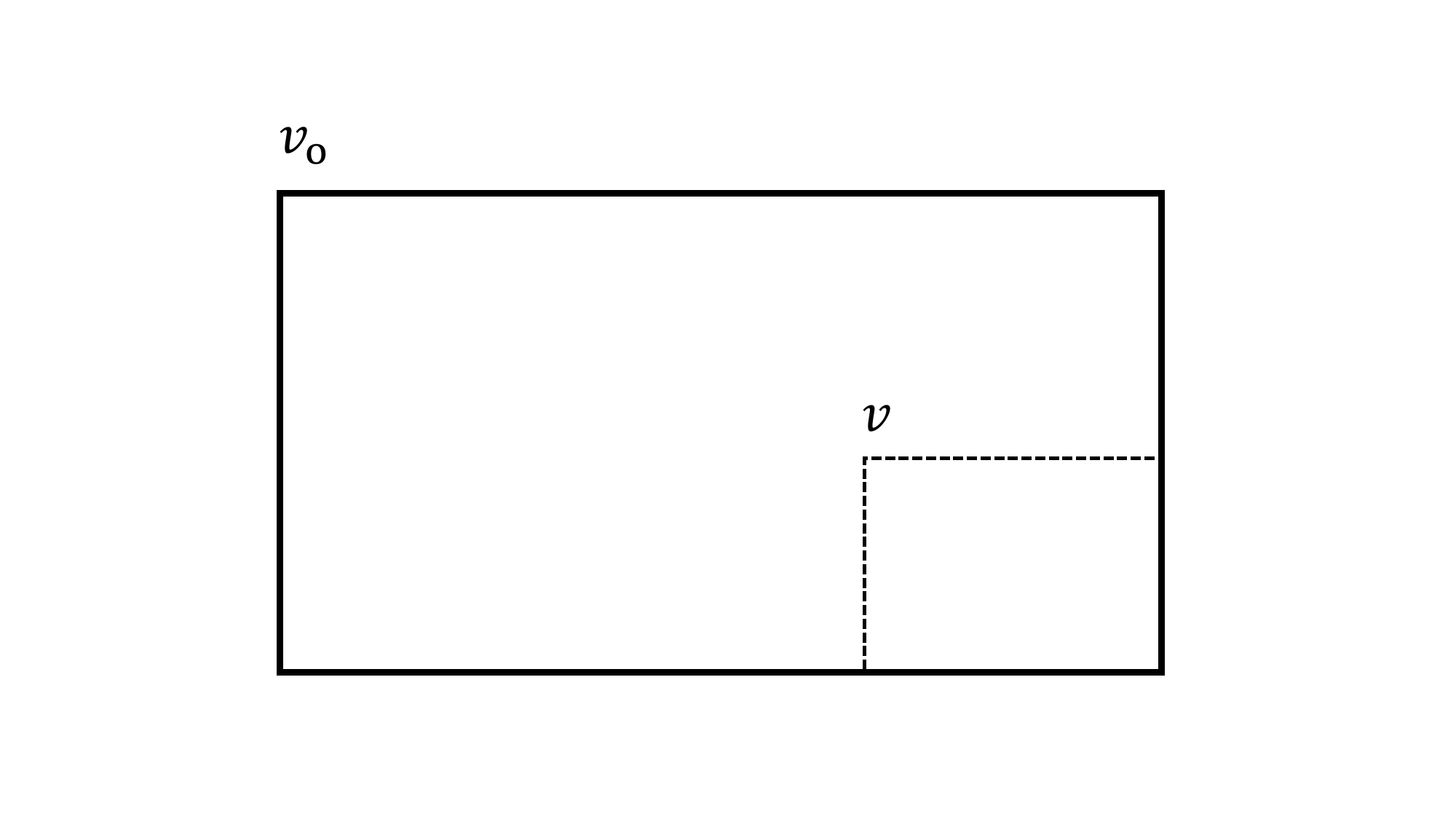}
	\caption{\small Diagram of the volumes considered by Einstein (this figure does not appear in his paper).}
	\label{fig:vol}
\end{figure}

Now, the difference of entropy for a change in volume (see Fig. \ref{fig:vol}) considering the frequency $\nu$ and the energy constant is:

\[ S - S_0 = \frac{E }{ \beta \nu} \ln\frac{v}{v_{\circ}}. \] 
Einstein notes that this formula shows that “the entropy of monochromatic radiation of sufficiently low density varies with the volume according to the same law as the entropy of an ideal gas or that of a dilute solution” (\cite{beck2}, p. 94).

In the following section, he proceeds to interpret this result using Boltzmann’s principle. He begins stating (\cite{beck2}, p. 94):

\begin{quote}
In calculating the entropy by molecular-theoretical methods, the word ``probability'' is often used in a sense that does not coincide with the definition of probability used in probability calculus. In particular, the ``cases of equal probability'' are often stated hypothetically when the theoretical models applied are sufficiently definite to permit a deduction instead of a hypothetical statement. 
\end{quote}
That is, it questions the meaning of $W$ in

$$S=\kappa \ln W,$$
where $\kappa$ is Boltzmann's constant. And continues:

\begin{quote}
I will show in a separate paper that, when dealing with thermal processes, it is completely sufficient to use the so-called ``statistical probability,'' and I hope that this will remove a logical difficulty that still hinders the implementation of Boltzmann's principle. Here, however, I shall only give its general formulation and its application to very special cases.
\end{quote}

He derives the functional form relating entropy to the probability of an instantaneous state, arriving at Boltzmann’s principle expressed as:

$$S-S_{\circ}=\frac{R}{N} \ln W,$$
where $W$ denotes the ``relative probability'' of a given state compared to the initial state, and $R/N$ is Boltzmann's constant (the gas constant divided by Avogadro's number). Einstein assumes that, for a system like an ideal gas, no point in space is privileged; therefore, the probability that the $n$ particles are found within a subvolume $v$ of the total volume $v_{\circ}$ is:
$$
W = \left(\frac{v}{v_{\circ}} \right)^n. 
$$
In his own words, this is the probability ``that at a randomly chosen instant of time all $n$ independently movable points in a given volume $v_{\circ}$ will be contained (by chance) in volume $v$'' (\cite{beck2}, p. 96).

In Section \textsection 6, Einstein interprets the expression for the entropy of radiation as a function of volume. Within Wien’s regime, the change in entropy due to a volume fluctuation is given by the equation:

\[ S - S_0 = \frac{R}{N} \ln{ \left( \frac{v}{v_0} \right)^{N E / R \beta \nu} }. \] 

Using Boltzmann's principle, it can be concluded that the probability that total radiation is in a subvolume $v$ of a box of total volume $v_{\circ}$ is: 

\begin{equation}  \label{eins1}
W = \left( \frac{v}{v_0} \right)^{N E / R \beta \nu}.   
\end{equation}
This expression leads to the conclusion that (\cite{beck2}, p. 97)

\begin{quote}
Monochromatic radiation of low density (within the validity of Wien's radiation formula) behaves thermodynamically as if it consisted of mutually independent energy quanta of magnitude $R\beta \nu/N$.
\end{quote}

\subsection{Analyses} \label{analy}

Let us now examine some of the analyses of Einstein’s argument that stand out within the vast literature on the history of quantum physics. Needless to say, we do not aim to provide a comprehensive review. Although the works we cite generally refer and assess one another, we believe that a detailed analysis and thorough comparison is as interesting as necessary to reach a well-founded conclusion.

Jon Dorling characterized Einstein’s argument as ``circular'' \citep{dorling}. According to him, it does not rely on an analogy. As we have seen, the core of Einstein’s reasoning lies in interpreting expression \eqref{eins1} as the relative probability that, at a given instant, all the radiation is confined within a partial volume $v$. According to Dorling, this interpretation inevitably leads to Einstein’s quantum hypothesis—namely, that radiation is spatially localized in discrete points carrying energy quanta of $h \nu$.

Let us now restate Dorling’s reasoning in our own words. He imagines the total volume divided into $n$ parts:

$$v_{\circ}=nv$$
($n$ is an integer). In general, the energy of the whole radiation can be expressed as:\footnote{The relation between constants in \eqref{eins1} is $R\beta/N=h$, where $h$ is Planck's constant.}
 $$E=x h \nu,$$
where $x$ is a real number. The relative probability in \eqref{eins1} can now be written as:

$$
\left(\frac{1}{n} \right)^{x}.
$$
Therefore, the probability of finding the energy in any of those $n$ subvolumes is:

\begin{equation} \label{dorl}
\left(\frac{1}{n} \right)^{x-1}.
\end{equation}

Dorling argues that the only coherent outcome for this approach is that $x$ must be a whole number. For example, if we consider an energy less than $h \nu$ (e.g., $x = 1/2$), the probability of finding the radiation in one half of the volume is:

$$
2^{1/2},
$$
which is greater than one. Since probabilities cannot exceed unity, this result is clearly unphysical. The same argument applies for any $x < 1$. Dorling concludes that “$E$ cannot lie between 0 and $h\nu$” (\cite{dorling}, p. 3)

If $E = h\nu$ (i.e., $x = 1$), the probability of finding the radiation in any subregion of the volume equals one. Since this subregion can be made arbitrarily small, the energy must be concentrated at a single point.

Finally, if the energy lies between $h\nu$ and $2h\nu$ (e.g., $x = 3/2$), then as seen from equation \eqref{dorl}, the probability of finding the energy in any subregion of the volume tends toward zero:

$$\frac{1}{n^{1/2}}.$$

Hence, in this case, the total energy cannot be concentrated at a single point, unlike what occurs when $x = 1$. Instead, the energy must be distributed across two or more regions, with some of these containing less than $h\nu$ of energy. At this point, Dorling appeals to his earlier conclusion—that energy less than $h\nu$ cannot exist at a single point—and therefore infers that the total energy cannot lie between $h\nu$ and $2h\nu$. In fact, this reasoning can be extended to any value of $x$ that is not an integer. For $x = 2$, the probability \eqref{dorl} is:

$$\frac{1}{n}.$$

Since this probability expression can be made arbitrarily small for large $n$, it follows that the energy cannot be concentrated at a single point. According to Dorling, and appealing again to his previous deductions, only the concentration of energy at two distinct points avoids internal contradiction. Arguments of this kind can be extended to any energy that is an integer multiple of $h\nu$, while for any non-integer value of $x$, the previous reasoning can be reiterated. In this way, Dorling argues that Einstein’s approach implicitly entails both the quantization and spatial localization of radiant energy.

However, we believe that Dorling’s analysis suffers from the same circularity he attributes to Einstein. Consider the first step of his argument: the failure to obtain physically reasonable probabilities for $E < h\nu$ leads him to conclude that such energies cannot exist. But one could just as plausibly infer that energies below $h\nu$ are possible, and that the expression in question simply cannot be interpreted as a relative probability. Still, in the case where $x$ is not an integer, one might conclude—contrary to Dorling—that energy cannot be concentrated at discrete points, but must instead be distributed across space. That is, it is not that Einstein imposed the quantum hypothesis from the outset; rather, it is Dorling who assumes that the expression must carry the same probabilistic meaning it holds in the context of an ideal gas. Under that assumption, it is unsurprising that expression \eqref{eins1} leads directly to the particle hypothesis.

In fact, it is Dorling himself who, in the appendix, admits that (\cite{dorling}, p. 7)

\begin{quote}
... it seems that the $W$ in Einstein's $S=\kappa \ln W$ cannot be \textit{generally} interpreted as a probability in the \textit{ordinary} sense since entropy differences are often well-defined for systems with different [\textit{sic.}] constants of motion (e.g. energy and particle number in the case of an ideal gas) while talk of the relative probability of systems with different constants of the motion does not seem to make sense (at any rate without some special explication which is nowhere given)
\end{quote}

Nevertheless, Dorling relies on his earlier analysis, which does not question this interpretation. That is precisely what we intend to do in the following. It is worth pausing to reflect on this point. If we understand it correctly, there is a mistake in this excerpt by Dorling. The first ``different'' (in ``for systems with different...'') should be ``the same''. That is, the probabilistic interpretation of entropy relies on defining a sample space characterized by the variables $E$, $v$, and $N$ in the case of ideal gases. For radiation, however, the variable $N$ has no meaning.

It is clear, then, that we must, in one way or another, confront the question of the nature of radiation and its behavior under fluctuations. Regarding Einstein’s argument, can all monochromatic energy be concentrated in an arbitrarily small subvolume? As both Dorling's and Einstein’s analyses suggest, if we assume that radiation consists of particles, the result aligns with what we know about ideal gases. F. E. Irons \citep{irons} emphasized this point, noting that Einstein implicitly assumed a specific behavior of radiation during a fluctuation—namely, a behavior that presupposes its corpuscular nature. In this respect, we concur with Irons in his critique of Dorling’s analysis (\cite{irons}, p. 272).

Once again, in Irons’s view, Einstein’s argument is ultimately tautological: assuming that radiation behaves like an ideal gas leads, unsurprisingly, to the conclusion that radiation behaves like an ideal gas. Irons discusses the case of adiabatic compression, where equilibrium states are connected. In such scenarios, the frequencies transform and vary in proportion to the energies associated with each mode of vibration:\footnote{By \textit{mode of vibration} we are refering to classical standing waves. Although Irons does not reference it, it is worth citing the paper in which Paul Ehrenfest first introduced the concept of adiabatic compression in the context of quantum theory and radiation laws: \citep{ehrenfest1911}. See also \citep{navarroperez2004}. We will return to it in Section \ref{inred}.}

\begin{equation} \label{adia}
\frac{E_{\nu}}{\nu}=\frac{E_{\nu_{\circ}}}{\nu_{\circ}}
\end{equation}
$$\nu^{3}v=\nu_{\circ}^{3}v_{\circ}.$$

These two adiabatically connected states would indeed have the same entropy—but they are not the states Einstein compares. In the scenario Einstein considers, the radiation retains the same frequencies throughout. In other words, the process in question is entirely unknown. To assume it as physically possible is, in effect, to impose a corpuscular behavior on radiation. In short, our limited understanding of the nature of radiation prevents us from validating a process like the one considered by Einstein. Moreover, Irons himself, once again in an appendix, echoes —quoting A. Brian Pippard— the absurdity of assigning an entropy to a fluctuation (\cite{irons}, p. 276):

\begin{quote}
...statistical fluctuations in volume are part of the nature of thermal equilibrium and `if we ascribe a definite value to the entropy of the gas in equilibrium we must ascribe it not to any particular, most probable set of configurations, but to the totality of configurations of which it is capable. Thus we see that the entropy... must be recognised as a property of the system and of its constraints, and that once these are fixed the entropy also is fixed'.
\end{quote}

We share Irons’s reservations regarding the very possibility of assigning an entropy to a fluctuation.

Another author who closely analyzed Einstein’s argument is John Norton, who famously described it as “miraculous” \citep{norton}. Norton highlighted Einstein’s boldness in identifying a signature of atomicity in radiation, noting a conceptual continuity with his earlier arguments—as well as those of other physicists and chemists—concerning Brownian motion, intermolecular forces, and related phenomena. In the case of radiation, where the number of molecules is not conserved, only a fluctuation of the type Einstein considered could reveal such a signature, suggesting spatial localization and independence. According to Norton, this is the primary merit of the argument.

However, Norton also points out that Einstein treats this ``rare fluctuation'' as if it were an equilibrium state (\cite{norton}, p. 86), and emphasizes the disanalogy between fluctuations in gases and in radiation. While the fluctuation Einstein considers is plausible in the context of gases—namely, the same energy distributed over unequal volumes at equal temperature—it is physically incoherent for radiation. Compressing radiation into a smaller volume would necessarily raise its temperature, making the analogy inapplicable.

Despite offering an engaging and insightful account, Norton appears to attribute more to Einstein’s argument than Einstein himself intended. At no point does Einstein explicitly address what is central to Norton’s reading: the non-conservation of the number of light quanta.

Anthony Duncan and Michel Janssen \citep{duncanjanssen2019,duncanjanssen2019w} also emphasize the disanalogy between gases and radiation. Like Norton, they note that Einstein applies equilibrium expressions to estimate entropies associated with non-equilibrium situations. Furthermore, they stress that the states Einstein compares cannot, in fact, share the same temperature—and they go on to calculate the temperature difference explicitly:

$$\frac{1}{\kappa T'}-\frac{1}{\kappa T}=\frac{1}{h \nu}\ln\left(\frac{v}{v_{\circ}} \right).$$

However, since Einstein only considers energy and volume, the temperature difference does not appear into the formal comparison he presents. To address this issue, Duncan and Janssen attempt to calculate the entropy change using ``...the number of ways $W$ of distributing $N$ energy packets over $G$ modes'' (\cite{duncanjanssen2019w}, p. 3). Yet, as far as we understand, their calculation is flawed. Although they aim to avoid introducing temperature, they ultimately do so, and in the final step—crucially—they equate the temperatures of the two states (\cite{duncanjanssen2019w}, p. 4). In our view, this renders their argument circular: they effectively calculate the same quantity twice by comparing entropies at different volumes and temperatures, but for the same energy.

When it comes to radiation, there is only one meaningful alternative: consider a fluctuation at constant temperature and differing energy. However, this scenario breaks the analogy with the gas, since the number of light quanta would then vary—something not accounted for in the gas model.

Duncan and Janssen also disagree with Dorling’s claim that Einstein’s argument demonstrates spatial localization. They argue that the volume dependence observed in the Wien regime arises not from quanta but from mode counting (\cite{duncanjanssen2019w}, p. 6).\footnote{They refer to subsequent developments by Pascual Jordan and others concerning wave-particle duality and the lack of localization. In our view, this duality does not apply in the Wien regime, but only when the full spectral range is considered. In the Wien limit, duality can be disregarded in favor of a purely corpuscular description.}

Let us now examine this last point by focusing on the notion of extensivity, using Planck’s entropy expressed in terms of the number of eigenmodes $n$ and energy quanta $p$, as formulated in a paper by Jean-Pierre Provost and Christian Bracco (\cite{provostbracco}, p. 1088):

\begin{equation} \label{entr}
S=\kappa \left[(n+p)\ln(n+p)-p\ln p - n\ln n) \right].
\end{equation}

These authors propose that, based on the symmetry of Planck’s entropy expression with respect to $n$ and $p$, it admits two possible interpretations: the canonical view, where quanta are distributed among resonators, and an alternative view, where ``coherence volumes'' of the radiation are distributed among quanta (\cite{provostbracco}, p. 1087). Within these volumes ``phases of different points are correlated''. Accordingly, they suggest interpreting “the associated number of complexions as a consequence either of a quantization of energy or of a quantization of volume” (\cite{provostbracco}, p. 1090). In the Wien regime, characterized by small coherence volumes, the quantization of volume would play no significant role. We do not discuss this proposal further here, as we believe it does not shed light on the specific issues we wish to address.

What interests us in expression \eqref{entr} is that it clearly reveals the connection to extensivity—particularly when written in the following form:

\begin{equation} \label{entr2}
S=\kappa \left[n\ln \left(1+\frac{p}{n}\right) + p\ln \left(1+\frac{n}{p}\right) \right].
\end{equation}

Is this expression extensive? Certainly, it is. However, note that extensivity appears differently in the two terms: proportionality with $n$ reflects extensivity with respect to volume, while proportionality with $p$ reflects extensivity with respect to energy. To recover the characteristic dependence of an ideal gas, the first term must be eliminated. Indeed, the case considered by Einstein corresponds to a low occupancy number, where $p << n$. Under this assumption, the entropy reduces to the form used by Einstein in \eqref{ewien}:

\begin{equation}  \label{entrw}
S \approx \kappa p \left[1 + \ln\left(\frac{n}{p} \right) \right].
\end{equation}

Where, then, does the volume dependence arise? It is contained in the number of modes, $n$. Extensivity is ensured by the linear dependence of entropy on $p$. Conversely, in the opposite limit—the Rayleigh-Jeans region—entropy scales with $n$, and the analogy with gases breaks down. We believe that this is precisely what Duncan and Janssen mean when they argue that the volume dependence is not related to the spatial localization of quanta, but rather to the low occupation number of the eigenmodes.

\subsection{Summary}

The main issues addressed regarding these analyses can be grouped as follows:

\begin{itemize}
\item \textit{The nature of radiation}. How does monochromatic radiation behave under a fluctuation of the kind proposed by Einstein?
\item \textit{The entropy of a fluctuation}. Is it meaningful to assign an entropy to a fluctuation?
\item \textit{Circularity of the argument}. Does Einstein presuppose the conclusion of his argument in its initial assumptions?
\item \textit{Scope of the analogy}. To what extent is the analogy between ideal gases and radiation valid?
\end{itemize}

The most crucial—and simultaneously most delicate—aspect of Einstein’s argument concerns the change in entropy. Specifically, Einstein compares two equilibrium entropy values that differ only by volume. But does it make sense to interpret one of these as representing the entropy of a fluctuation? As we have seen, numerous authors have raised doubts about this point. Indeed, it is conceptually problematic to speak of the \textit{entropy of a fluctuation}. By definition, entropy accounts for all possible microscopic configurations, including those that contribute to what we call fluctuations. Moreover, from a thermodynamic standpoint, fluctuations do not formally exist, and thus thermodynamic expressions cannot meaningfully be applied to describe them. From a statistical perspective, one must appeal to Boltzmann’s principle—as Einstein does—or to the canonical ensemble.\footnote{In the canonical ensemble, fluctuations occur in energy at constant temperature, which is not the case Einstein considers.} For ideal gases, this approach is viable because the number of microstates (complexions) can be computed; however, for radiation, such a computation is not possible. What Einstein does, instead, is reverse the reasoning: rather than deriving entropy from probability, he derives the probability $W$ from the entropy $S$. Since $S$ is defined for equilibrium states, the resulting $W$ must also correspond to an equilibrium configuration. In our view, Einstein’s procedure is formally sound, but it should be understood strictly as a comparison between two equilibrium entropies, not as a description of a genuine fluctuation.

It is understandable that Einstein's argument can be interpreted, as some authors have done, as imposing ideal gas behavior on radiation, which is ultimately what it is all about. We have seen that Einstein, following Planck and others, argues that the entropy of radiation can be expressed as the sum of monochromatic entropies \eqref{entrsum}. No analogous decomposition is possible for an ideal gas, due to the presence of an additional constraint: the number of particles, $N$, a variable that has no meaningful analogue in the case of radiation. To preserve the analogy, one would have to compare different ideal gases—e.g., gases composed of particles with varying masses—to radiation with different frequencies. In that framework, the entropies would indeed be additive. However, this highlights why the same procedure used for radiation \eqref{lagr} cannot be directly applied to gases.

This discrepancy also explains why an ideal gas can undergo a volume change while keeping both its energy and temperature constant—since these two quantities are proportional for ideal gases—whereas for radiation, such a process is not feasible. Any increase in energy density necessarily entails a change in temperature. As we have noted, one is forced to choose between two incompatible scenarios: an energy fluctuation at constant temperature (as in the canonical ensemble, because temperature is a non-mechanical variable), or, alternatively, a constant energy fluctuation (as considered by Einstein) that involves a variation in temperature. Yet, if the situation is genuinely a fluctuation, temperature ceases to be well-defined.

In sum, to what extent does Einstein's analogy make sense? Regarding the possibility of a fluctuation of the type proposed by Einstein, it clearly makes sense with ideal gases because we know their behavior out of equilibrium. Now, an ideal gas is a very particular system with very peculiar properties, such as the fact that its energy does not depend on volume. For radiation, in general, it does not make sense to propose such behavior. This suggests the circularity pointed out by various authors. Hence, the simplicity of the ideal gas system, in this sense, is misleading.

\section{Einstein and Boltzmann's principle} \label{boltzmannp}

Einstein himself expressed doubts about his argument. In this section, we will show that these doubts ran parallel to those surrounding Boltzmann's principle. The gradual acceptance and consolidation of kinetic theory \citep{smithseth,taschetto}—an area to which Einstein had made significant contributions through his study of Brownian motion—was, in Einstein’s case, accompanied by a growing unease with Boltzmann’s principle, at least in its application to radiation. Although this process culminated in Satyendra Nath Bose’s 1924 work and Einstein’s subsequent extension of it, our analysis will focus on the earlier phase, roughly between 1905 and 1914, when Einstein turned to the adiabatic hypothesis in an effort to justify Boltzmann’s principle itself.

Einstein was the first to coin the expression ``Boltzmann’s principle'' in the very paper under discussion, where he also provided a derivation—or at least a justification—of this principle.\footnote{Some of the ideas presented in this section were previously developed in \citep{navarroperez2002}.} As several historians have pointed out \citep{norton, duncanjanssen2019}, it is striking that in 1905 Einstein did not draw upon the results he had previously obtained in his foundational works on statistical mechanics. Why is this the case?

In his so-called statistical trilogy \citep{einstein1902, einstein1903, einstein1904}, Einstein had indeed emphasized the tendency of physical systems to evolve from less probable to more probable configurations. Notably, in his 1903 paper, he also derived the following expression:

\begin{equation} \label{mikr}
S = 2k \ln{\omega(E)},
\end{equation}
where $\omega(E) \, dE$ represents the volume of the system’s phase space. Einstein claimed that this expression is “completely analogous to the expression found by Boltzmann for ideal gases and assumed by Planck in his theory of radiation” (\cite{beck2}, p. 68). However, as Jos Uffink has pointed out (\cite{uffink}, p. 54), the expression \eqref{mikr} is not, in fact, fully analogous to Boltzmann’s entropy, which is applicable even outside equilibrium. This discrepancy may partly explain why Einstein chose, in 1905, to re-derive and justify Boltzmann’s principle independently of the statistical mechanics he had previously developed. It is plausible that by that time, Einstein no longer regarded the 1903 expression as fully equivalent to Boltzmann’s formulation.

In those earlier papers, we do encounter the concept of fluctuation. Einstein, for instance, calculated the energy fluctuations of a system held at fixed temperature:

\begin{equation} \label{typical}
\langle \epsilon^{2} \rangle =2 \kappa T^{2} \frac{\langle dE \rangle}{dT}.
\end{equation}
This expression had also been previously derived by Gibbs (\cite{gibbs}, p. 72). It is well known that, in the subsequent section, Einstein refers to radiation as the “kind of physical system for which we can surmise from experience that it possesses energy fluctuation” (\cite{beck2}, p. 76). The precise nature of the empirical evidence he had in mind remains unclear \citep{rynorenn}.

Nonetheless, as early as 1904, Einstein recognized that the applicability of his statistical results to radiation was far from guaranteed. As he noted (\cite{beck2}, p. 76):

\begin{quote}
Of course, on purpose, we are not permitted to assert that a radiation space should be viewed as a system of the kind we have assumed, not even if the applicability of the general molecular theory is granted. Perhaps one would have to assume, for example, that the boundaries of space vary with its electromagnetic states. However, these circumstances need not be considered, as we are dealing with orders of magnitude only.
\end{quote}

Indeed, in 1905 Einstein expressed criticism of the concept of ``probability'' as employed in “molecular-theoretical” methods, arguing that “the cases of equal probability” are often introduced hypothetically, even in situations where the underlying theoretical model would permit a deductive treatment (\cite{beck2}, p. 94). In response, he proposed a notion of ``statistical probability,'' essentially an inversion of Boltzmann’s principle, whereby probability is inferred from entropy rather than the other way around. To justify this redefinition, he appealed to a familiar ``theoretical model'': the ideal gas. In the case of radiation, however, Einstein did not possess an analogous theoretical framework that would allow him to compute the statistical weight (or fluctuation) directly. His inversion thus served as a workaround, offering a way to define probability in the absence of a microscopic model. This move can plausibly be read as a critique of Boltzmann’s and Planck’s reliance on counting complexions, as Abraham Pais suggested (\cite{pais}, p. 72). Einstein appears to favor deriving statistical relations from thermodynamic principles rather than through combinatorial methods.

In a 1907 paper, Einstein reaffirmed his confidence in Boltzmann’s principle, writing (\cite{einstein1907}, p. 225):

\begin{quotation}
At first glance the theoretical examination of the statistical law that governs these fluctuations would seem to require that certain stipulations regarding the molecular model must be applied. However, this is not the case. Rather, essentially it is sufficient to apply the well-known Boltzmann relation connecting the entropy $S$ with the statistical probability of a state.
\end{quotation}

As is well known, in 1909 Einstein introduced a new treatment of radiation based on fluctuations, markedly different from the approach he had taken in 1905, though still grounded in the same inversion of magnitudes \citep{einstein1909, einstein1909b}. In this later work, the fluctuations he considered were more conventional: he analyzed energy fluctuations at fixed temperature. Moreover, he restricted his attention to small deviations from equilibrium, in contrast to the more drastic hypothetical fluctuations considered in 1905. Once again, Einstein criticized the absence of a rigorous definition of probability, this time explicitly referencing both Boltzmann and Planck (\cite{beck2}, p. 362). He proposed a definition consistent with his earlier 1903 formulation and reiterated his full confidence in Boltzmann’s principle, stating: “A theory yielding values for the probability of a state that differ from those obtained in this way must obviously be rejected” (\cite{beck2}, p. 364).

In the absence of a microscopic model capable of calculating the number of complexions, Einstein, as in 1905, derived probabilities from entropy. He considered volumes $V>>v$, ``two interconnected spaces bounded by diffusely, completely reflecting walls'' (\cite{beck2}, p. 364). Einstein calculates the total entropy in the vicinity of equilibrium using a Taylor expansion ($\Sigma$ and $\sigma$ are the corresponding entropies for $V$ and $v$):

\begin{equation}
S=\Sigma + \sigma=const+\frac{1}{2} \left\{\frac{d^{2} \sigma}{d \epsilon^{2}} \right\}_{\circ} \epsilon^{2} + \cdots,
\end{equation}
where $\epsilon$ is energy. The value of the average energy  of these fluctuations can be calculated with this expression and Boltzmann's principle, with:

\begin{equation}
dW= const. e^{\frac{1}{2}\frac{N}{R} \left|\frac{d^{2} \sigma}{d \epsilon^{2}} \right|_{\circ}\epsilon^{2}} d\epsilon
\end{equation}

In this case, the problems we discussed in the 1905 fluctuation disappear: the temperature is constant, and therefore, energy can fluctuate. Moreover, the formula he obtains coincides with the one he obtained in 1904:

\begin{equation}
\overline{\epsilon^{2}}=\frac{1}{\frac{N}{R} \left\{\frac{d^{2} \sigma}{d^{2} \epsilon} \right\}_{\circ}}.
\end{equation}
Duncan and Janssen note that Einstein avoids using the standard formula of statistical mechanics \eqref{typical} because it is not applicable to radiation (\cite{duncanjanssen2019}, p. 102). However, in accordance with our claims, we believe the issue is not whether the formula applies to radiation, but rather that Einstein’s approach had shifted during those years: his focus turned to the interpretation of Boltzmann’s principle. 

Substituting the entropy associated with Planck’s law yields:

\begin{equation} \label{fluctu}
\overline{\epsilon^{2}}=\frac{R}{N \kappa}\left\{\nu h \eta_{\circ} + \frac{c^{3}}{8 \pi \nu^{2}d\nu}\cdot \frac{{\eta_\circ}^2}{\nu}  \right\},
\end{equation}
where $\eta_{\circ}$ denotes the energy. As has been said, Einstein argued that each of these terms could be associated with the corpuscular or wave manifestations of radiation. In fact, this is also the paper in which Einstein introduces the now-famous \textit{Gedankenexperiment} through which he derives the wave–corpuscle duality from Planck’s law. The setup consists of a mirror that is transparent to all frequencies except those in the interval ($\nu$, $\nu + d\nu$), and which moves perpendicularly to its surface. The mirror is placed inside a cavity filled with electromagnetic radiation and gas, all in thermal equilibrium. The momentum fluctuation originated by the fluctuation of radiation pressure consists of two terms:

$$
\overline{{\Delta}^2} = \frac{1}{c} \Big[ h \rho \nu + \frac{c^3}{8\pi } \frac{\rho^2}{\nu^2}  \Big] d\nu f \tau ,$$
being $\tau$ a small time interval and $f$ the surface of the mirror. In such a context, Einstein suggested for the first time the dual constitution of light (\cite{beck2}, p. 379):\footnote{This excerpt is not from the same paper, but from the lecture he delivered in the ``Session of the Division of Physics of the 81st Meeting of German Scientists and Physicians in Salzburg on September 21, 1909'' (\cite{beck2}, p. 379)} 

\begin{quote}
It is even undeniable that there is an extensive group of facts concerning radiation that shows that light possesses certain fundamental properties that can be understood far more readily from the standpoint of the wave theory. It is therefore my opinion that the next stage in the development of theoretical physics will bring us a theory of light that can be understood as a kind of fusion of the wave and emission theories of light.
\end{quote}

It is important to note that while in 1905 Einstein relied solely on Wien’s law—which is valid only at high frequencies—and derived only the corpuscular term or particle-like behavior of light, in 1909 he employed Planck’s law—which holds across the entire spectrum—and thus obtained both the corpuscular and wave-like aspects of light. Furthermore, in 1905 he used Boltzmann’s principle to compute the entropy difference between two equilibrium states with the same energy, thereby determining the relative probability between those states. In contrast, in 1909 both contributions arise in the context of calculating energy fluctuations at fixed temperature.

In 1910, in a paper on critical opalescence, Einstein includes a section titled ``General Remarks about Boltzmann’s Principle.'' There, he underscores the difficulty of assigning meaning to $W$, and expresses skepticism about the principle itself (\cite{beck3}, p. 232):\footnote{Emphasis in the original.}

\begin{quote}
$W$ is commonly equated with the number of different possible ways (complexions) in which the state considered --which is completely defined in the sense of a molecular theory by observable parameters of a system-- can conceivably be realized. In order to be able to calculate $W$, one needs a \textit{complete} theory (perhaps a complete molecular-mechanical theory) of the system under consideration. Given this kind of approach, it therefore seems questionable whether Boltzmann’s principle \textit{by itself} has any meaning whatsoever, i.e., without a \textit{complete} molecular-mechanical or other theory that completely represents the elementary processes (elementary theory). If not supplemented by an elementary theory or --to put it differently-- considered from a phenomenological point of view, equation (1) [Boltzmann’s principle] appears devoid of content.
\end{quote}

Einstein presents his definition of probability based on the fraction of time a system spends in each state. Since one of these states is overwhelmingly more probable than the others—differing by orders of magnitude—the system’s evolution will almost always tend toward that state. Moreover, Einstein argues that it is meaningless to speak of the probability of a single state, as such a state has measure zero; instead, one must always refer to a finite region, no matter how small. We interpret this as a clear departure from the concept of complexion.

In the second section of the paper, Einstein again inverts Boltzmann’s principle to analyze small deviations from equilibrium. He further asserts that for “deviations from thermodynamic equilibrium as small as those considered in our case, the quantity $S - S_{\circ}$ has an intuitive meaning” (\cite{beck3}, p. 235 ). Without providing a concrete example, he argues that whenever an equilibrium state can be reversibly connected to a nearby non-equilibrium state, the use of the version of Boltzmann’s principle he proposes is justified.

Let us note two points. First, Einstein is explicitly attempting to extend the application of Boltzmann’s principle to non-equilibrium states. While Boltzmann had already suggested this possibility, Einstein did not yet know how to render it operational beyond the case of ideal gases. Second, this line of reasoning anticipates Ehrenfest’s adiabatic hypothesis—a concept Einstein himself would later name and adopt within a few years \citep{einstein1914,navarroperez2006}. These uncertainties are expressed most clearly in a letter he wrote to Besso in the summer of 1911:\footnote{Einstein to Besso, Second half of 1911. In \cite{beck5}, p. 197}
\begin{quote}
I am still scribbling about the Boltzmann's principle... Unfortunately, we lack the knowledge required to determine the entropy in a general way while taking into consideration the statistical irregularities, so that we must content ourselves with an approximation. It is also embarrassing that we can only speak of the entropy of such states that are at least in principle thermodynamically realizable. It is thus always necessary to contrive an analogue of the semipermeable membrane, which, however, in most cases cannot be accomplished. We cannot conceive of ``walls'' that will keep the thermal energy of a subsystem within specific limits. For that reason, the entropy of a heat content region cannot be determined either, so that the calculation of the temperature fluctuation lacks rigor. The situation is just as bad when it comes to the energy of a periodically oscillating system. Some means ought to be found for extending the entropy concept of thermodynamics to cases that cannot be regarded as instances of thermodynamic equilibrium.
\end{quote}

By ``oscillating systems,'' we understand Einstein to be referring to radiation. In other words, he is beginning to suspect that, beyond the ideal gas, Boltzmann’s relation is more of a formal device than a physically meaningful principle. Although he does not question its validity, it proves of limited practical use out of equilibrium: to determine the entropy of a given state, it must be adiabatically connected to an equilibrium state. Therefore, there is no general method to calculate entropy for non-equilibrium states. Einstein also shared these reflections during the discussion that followed his presentation at the Solvay Conference (\cite{solvay}, p. 436). After reaffirming his confidence in the principle of energy conservation, he once again referred—this time publicly—to oscillatory systems (\cite{beck3}, p. 426):

\begin{quote}
In my opinion, another principle whose validity we must maintain unconditionally is Boltzmann’s definition of entropy through probability. It is this principle that we owe the faint glimmer of theoretical light we now see shed over the question of states of statistical equilibrium in processes of an oscillatory character. But there is still the greatest diversity of opinion as regards the content and domain of validity of this principle. 
\end{quote}

\begin{figure} [ht]
	\centering
		\includegraphics[trim = 5mm 60mm 10mm 50mm, clip,width=13cm]{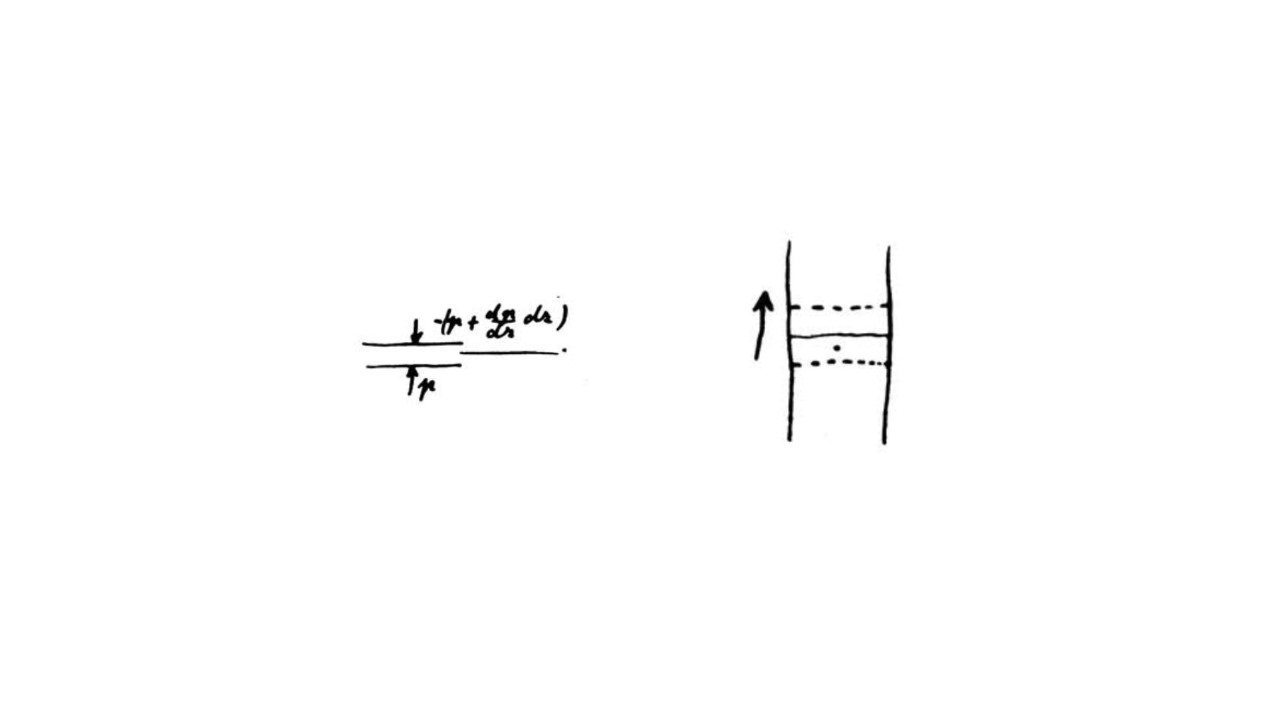}
	\caption{\small Drawings of Einstein of the system of a suspended particle. In \cite{kleinkox}, p. 234}
	\label{fig:suspended}
\end{figure}

 It is interesting to observe how Boltzmann’s principle serves as a guiding idea for Einstein, even though it has yet to be firmly established. He introduces his definition of probability in terms of time intervals and states general properties of the entropy for systems composed of subsystems:
\begin{equation}
 S = \sum S \qquad W = \prod W,
 \end{equation}
 which, however, are generally not valid for radiation. He also justifies the inversion of the procedure and provides a concrete example for calculating non-equilibrium entropy in the case of a massive particle immersed in a fluid—one that does not remain in its equilibrium position due to fluctuations but whose different states can be connected mechanically (see fig. \ref{fig:suspended}). Both in his correspondence with Besso and during the Solvay Conference, Einstein employed this example to clarify his relationship with Boltzmann’s principle and to support his 1905 argument. In order to compare probabilities via Boltzmann’s principle, the energy must remain constant. For this reason, Einstein placed the system in a thermal bath: any work done on the particle, changing its potential energy, must be compensated by an equivalent energy exchange. In other words, these transformations are not adiabatic in the Ehrenfest sense, where the energy varies. Similarly, in the 1905 fluctuation, the energy also remained constant. However, unlike the suspended particle, we lack any intuitive way to visualize this process.

 In the discussion, Lorentz had to explain to Poincaré that Einstein was not following Gibbs’s statistical approach. Einstein, for his part, insisted on his proposal to demonstrate the usefulness of Boltzmann’s principle, even without a theoretical framework capable of calculating probabilities. His focus was clearly on radiation. Neither Wien nor Planck believed the principle could be directly applied to radiation. In this context, Einstein revisited his 1905 argument, highlighting the importance of fluctuations —particularly in the Wien regime- where radiation behaves in a way that contradicts its wave-like nature and instead resembles an ideal gas. Once again, Einstein uses equilibrium entropy to calculate the entropy associated with a fluctuation.

 What interests us here is that Einstein proposes yet another inversion—this time, a more profound one: it is quantum theory that will ultimately enable us to understand Boltzmann’s principle (\cite{beck3}, p. 431):

\begin{quote}
The quantum hypothesis is a provisional attempt to interpret the expression for the statistical probability $W$ of the radiation. By conceiving radiation as consisting of small complexes of energy $h \nu$, one found an intuitive interpretation of the probability law for low-intensity radiation. I emphasize the provisional character of this auxiliary idea, which does not seem compatible with the experimentally verified conclusions of the wave theory.
\end{quote}

It is hard not to see here an intuition of the future developments that would lead to quantum statistics and the concept of indistinguishability. But it was not until 1924, with Bose's work, that events in this direction began. 

Lorentz—likely more familiar with the subject than the other participants, as Einstein had given a lecture on fluctuations in Leiden in February 1911 (\cite{beck3}, p. 450)—defended Einstein’s calculation. After all, even though the system was out of equilibrium, the entropy of the radiation involved in the well-known 1905 fluctuation should be calculable in the manner Einstein proposed. Planck, however, disagreed.

We conclude our analysis here, though not without noting that shortly afterward, Ehrenfest’s adiabatic hypothesis began to influence Einstein’s thinking. In fact, it was introduced at the Solvay Conference through the example of a pendulum whose length is slowly varied. We have analyzed elsewhere the emergence of this hypothesis in detail, as well as his influence on Einstein  \citep{navarroperez2004, navarroperez2006}. What is worth emphasizing here is that, in 1914, Einstein published a paper in which he carried out yet another inversion (the third!): rather than starting from Boltzmann’s principle, he treated it as a consequence of the adiabatic hypothesis—contrary to Ehrenfest’s own view. The confidence Einstein had expressed in 1911 was, by then, beginning to fade.

Indeed, Ehrenfest had to clarify in a letter to Einstein that, for years, he had been using statistical weight functions that departed from the original spirit of Boltzmann’s formulation:\footnote{Ehrenfest to Einstein, 21 May 1914. In \cite{hentscheleins}, p. 20. The \textit{statistical weight function} defines the weights on phase space (see equation \eqref{wf}). In the classical case, it would be uniform. In the quantum case, it would be a discrete function.}

\begin{quote}
a) Planck, Einstein, Debye work with $G(q,p,\underline{a})$, therefore it is worthwhile to examine why these people do come up with $\mu \delta Q= \delta W$   with such anti-Boltzmann spirited $G$’s.

b) Only \textit{once} did anyone work with $G(q,p,a,\underline{T})$: Herzfeld. This displeases me.

c) Ideal gas can ``of its own accord'' happen to shrink to half its volume on one occasion and to a $1/3$ on another, as if labeled item (a) had compelled it --Classical Hertzian resonators at frequency $\nu_{\circ}$ could ``\textit{by coincidence}'' all be ``stunned'' at once. On arrival at the Planck ellipses they will belong in frequency $\nu(\alpha_1, \beta_{1})$, in another instance \textit{by chance} to Planck ellipses that belong in frequency $\nu(\alpha_2, \beta_{2})$ --as if they had been \textit{pressed} on to these ellipses through corresponding $\alpha$, $\beta$ values with the aid of the quantum hypothesis lever.--\textit{Calculate the quotients of the probabilities of both these coincidences.}

Yes sir--\textit{this} would be the entropy calculation in Boltzmann’s spirit.

You see I \textit{understand} your comment.

But did Planck, \textit{you}, and Debye calculate \textit{it like this}?--No!-- Rather with $G(q,p,a)$  see e.g., Einstein, Ann. D. Phys. 22 (1907) p. 182 bottom. 
\end{quote}

Ehrenfest was referring to Einstein’s paper on specific heats \citep{einstein1907sh}. In that work, Einstein indeed employed a statistical weight function, rather than invoking Boltzmann’s principle. In fact, if Einstein ever performed a calculation truly in the spirit of Boltzmann’s principle, it was in 1905, in his analysis of the ideal gas.

\subsection{Still a fluctuation?}

After what we have seen, we cannot help but wonder whether Einstein continued to consider the process considered in 1905 as a fluctuation at the same level as that of 1909. We have seen that the latter is a typical energy fluctuation with a fixed temperature. The 1905 fluctuation is more of a temperature fluctuation with the same energy. Although we cannot be categorical, we do think that Einstein slightly modified his own explanation  when he presented his argument. In the notes from the lectures he gave in Leiden in February 1911, to which we referred earlier, we find the term \textit{Schwankung der Strahlungsenergie} for energy fluctuations (\cite{kleinkox}, p. 450). The 1905 argument is developed earlier, but apparently not explicitly associated with fluctuations.

In the presentation he gave at Solvay, Einstein introduced a wall separating the two cavities with a hole in the middle (\cite{beck3}, p.430-431).\footnote{He had given a similar presentation of the argument in Neuchâtel the previous year, 15 May 1910, at the meeting of the \textit{Société Suisse de Physique in Neuchâtel} \citep{einstein1910b}.} He claims that it is ``an analysis analogous to the one just brought forth for the case of a suspended particle.'' In other words, he attempts to justify the calculation of entropy in a fluctuation through a reversible transformation. In the radiation system he presents, the distribution of energy will be proportional to the volumes, but 

\begin{quote}
...because of the irregularities if the radiation process, all other distributions compatible with the given value $E$ of the total energy will also occur. To each of the distributions ($E_1, E_2$) belongs to probability $W$. But to each of the distributions there belongs also a definite value of the entropy $S$. $S$ and $W$ must be connected by Boltzmann’s equation. Since the entropy of any distribution of this kind can be found from the radiation law, one obtains the statistical probability $W$ of any distribution from Boltzmann’s equation.
\end{quote}

By 1911, it becomes clearer that Einstein considered, in the case of radiation and in the absence of a kinetic-molecular theory, that the probability can only be computed using the entropy of equilibrium states. Furthermore, the singularity of the process under consideration is reduced by the fact that Einstein’s argument involves different proportions of radiation distributed between the two subvolumes, rather than the extreme case where one cavity is left entirely without energy.

\begin{figure} [ht]
	\centering
		\includegraphics[width=6cm]{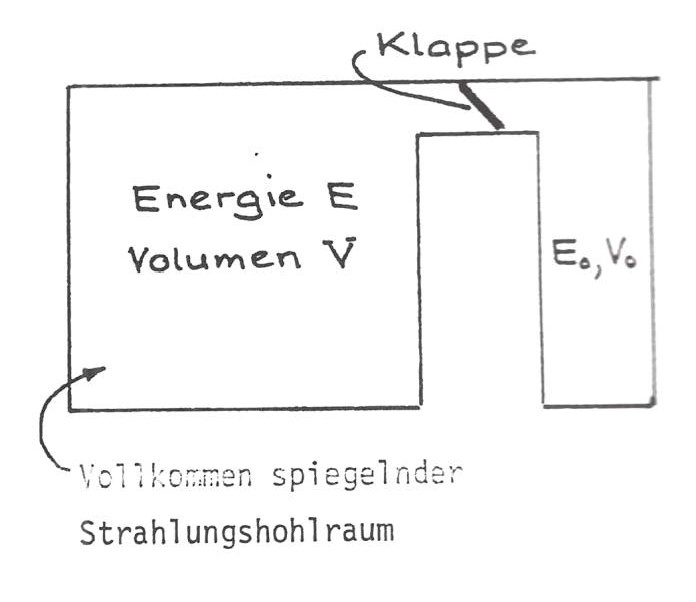}
	\caption{\small Drawing that appears in Dällenbach's notes. See footnote \ref{d}}.
	\label{fig:dalle}
\end{figure}

We find a similar presentation in Walter Dällenbach’s notes from the 1913 summer semester of the \textit{Statistische Mechanik} course taught by Einstein in Zurich.\footnote{Statistische Mechanik. Prof. Einstein. Vorlesungsnachschrift von Walter Dällenbach. Sommersemester 1913 an der ETH Zürich. In Helvetic Archives, SLA-Einstein-SSLB-C-2-08.\label{d}} Quantum theory appears near the end of these notes. There, we find the 1905 argument, but now with a valve (\textit{Klappe}) separating volume $v$ from $v_{\circ}$ (see Fig. \ref{fig:dalle}). Dällenbach explicitly writes that “$S_{\circ}$ is the entropy when all the radiation is located at one moment completely on one side of the volume.” This can be interpreted as a fluctuation, but the presence of the valve invites us once more to understand the argument instead as a comparison between equilibrium states.

In summary, the ambiguity between equilibrium states and fluctuations persists—and arguably becomes even more pronounced—over the years.


\section{Are there lightquanta beyond the infrared spectra?} \label{inred}

Everything we have said so far refers to the low-density regime, where Wien's law of radiation can be used instead of Planck's. Physics has changed considerably since that distant 1905, but as we know, light quanta, transmuted into photons, have survived to this day. Needless to say, with characteristics that make them completely different from what Einstein and his contemporaries imagined. Einstein was never satisfied with either his original proposal or the forms it subsequently took. However, there is no doubt that Einstein himself planted the first seed. Therefore, it is pertinent to ask the question: Does it make sense to extend Einstein's hypothesis beyond Wien's range, where it was born nurtured by statistical physics?

We will address this question from two perspectives: one historical and the other by referring to current theories. From a historical perspective, we will see how Einstein himself (and others) found arguments that reinforced the idea that it was in the low-density regime that the corpuscular properties of radiation showed up. Quantum field theory (QFT), which emerged in the late 1920s, forces us to reframe the question: in what range does it make sense to perform a classical (that is, undulatory) approximation of electromagnetic radiation?

Let us begin by reviewing combinatorics, since Planck's formula \eqref{entr2} allows to analyze its behavior at both ends of the spectrum. We saw that in the Wien limit, \eqref{entrw}, extensivity is guaranteed by the dependence on $p$, the number of quanta. If we now look at the other limit, that of high occupation ($p>>n$), entropy becomes:

\begin{equation}  \label{comx}
S=\kappa n \left[1 + \ln\left(\frac{p}{n} \right) \right].
\end{equation}
This expression, where only the extensivity resulting from volume survives, is the truly wave-like one and corresponds to the Rayleigh-Jeans law.

It is tempting to turn to combinatorics to characterize both limits, as has been done in other occasions \citep{darrigol}. In the Planck case, the combinatorial formula is:

$$W=\frac{(n+p-1)!}{p!(n-1)!}.$$
In Wien's limit, we can assume that there is at most one quanta in each mode, that is:

$$W=\frac{n!}{(n-p)!p!},$$
which in this limit coincides with distributing $p$ quanta in $n$ ways and subtract the equivalent ones:
$$W=\frac{n^{p}}{p!}.$$
These two expressions lead to the same result, expression \eqref{entrw}. This analysis was presented by Ehrenfest and Kamerlingh-Onnes in a paper in which they also published a derivation of Planck's combinatorial formula, widely used today \citep{ehrkam}. Their conclusion was clear (\cite{klein}, p. 356):\footnote{Emphasis on the original.}

\begin{quote}
\textsc{Planck} does not deal with really mutually free quanta $\epsilon$, the resolution of multiples of $\epsilon$ into separate elements $\epsilon$, which is essential in his method, and the introduction of these separate elements has to be taken ``cum grano salis'', it is simply a formal device entirely analogous to our permutation of the elements $\epsilon$ or $0$.
\end{quote}

 In contrast, Einstein's quantization did imply a certain individuality of the energy quanta (\cite{ehrkam}, p. 355):\footnote{Emphasis on the original.}  

\begin{quote}
\textsc{Einstein} really considers $P$ similar quanta, existing \textit{independently of each other}. He discusses for instance the case, that they distribute themselves irreversibily from a space of $N_{1}$ cm$^{3}$ over a a larger space of $N_{2}$ cm$^{3}$ and he finds using Boltzmann's entropy-formula that this produces a gain of entropy...
\end{quote}
Note how Ehrenfest --a personal friend of Einstein and very knowledgeable in statistical methods-- and Kamerlingh-Onnes describe the 1905 fluctuation: they prefer to describe the process in the reverse time sense: an initial state in which we have the radiation confined in a small volume that, irreversibly, expands to occupy a larger volume. This is not a fluctuation but a free expansion. In the case of ideal gases, temperature does not change. For radiation, it does.

\begin{figure} [ht]
	\centering
		\includegraphics[trim = 45mm 70mm 10mm 30mm, clip,width=15cm]{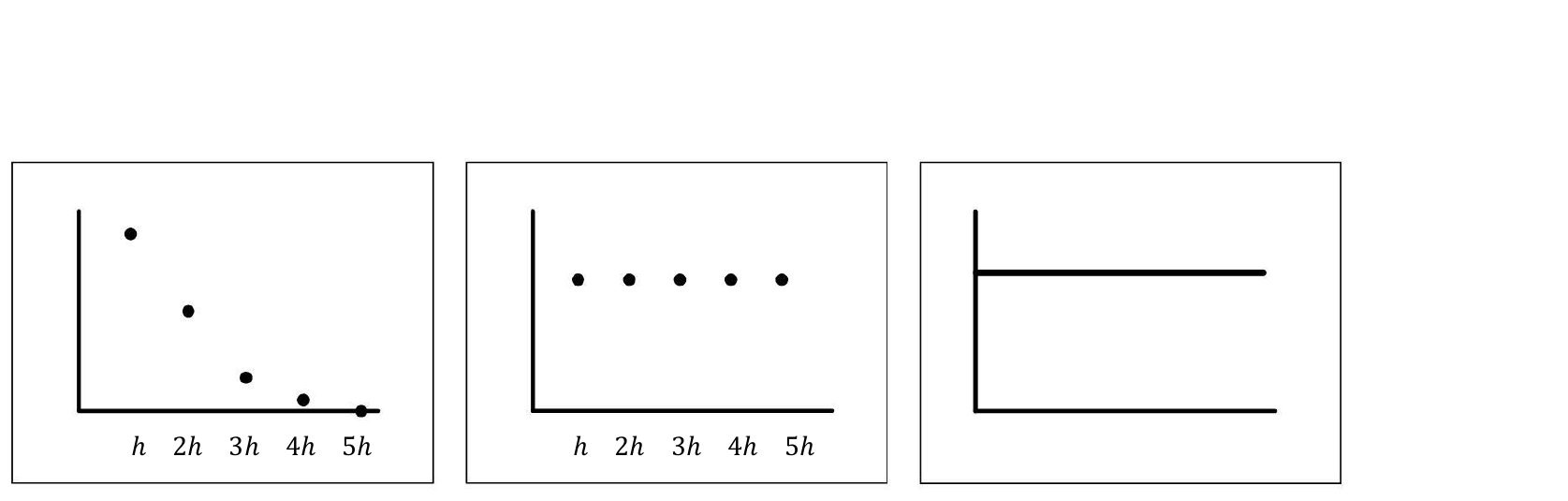}
	\caption{\small Schematic dependence of statistical weights $G(E_{\nu}, \nu)$ corresponding to the Wien, Planck and Rayleigh-Jeans laws.}
	\label{fig:ehr11}
\end{figure}

In fact, Ehrenfest had presented a brilliant paper in 1911, in which he exhaustively analyzed the blackbody problem \cite{ehrenfest1911}. We will not go into detail about it, as we have done so elsewhere \citep{navarroperez2004}, but we do want to note that in that paper, Ehrenfest characterized the statistical difference between the quantizations of Planck and Wien. He defined the statistical weight function $G(E_{\nu}/\nu)$ as follows:

\begin{equation}  \label{wf}
    \gamma(E_{\nu}, \nu)=Q(\nu)G(E_{\nu}/\nu).
\end{equation}
This way of writing the statistical weight function comes from applying an adiabatic transformation \eqref{adia}. For the Rayleigh-Jeans law we have $G(E_{\nu}/\nu)=const.$, and for Planck and Wien $G(E_{\nu}/\nu)$ is a discrete function with non-zero values in $E_{\nu}/\nu=nh$ (see fig. \ref{fig:ehr11}). In the case of Planck each value of $E_{\nu}/\nu$ has the same statistical weight, and in the Wien's case it decreases as $n!$ While classical radiation law corresponds to a uniform distribution, Planck and Wien allow values proportional to $\nu$, but with different weights.

Ehrenfest showed that quantization was inescapable, given the empirical form of the radiation law. While the shape of the radiation law in the infrared implied a certain discontinuity at the origin, the quantization of the energy came from the behavior in the violet range, as Einstein argued in 1905 and 1909.

Certainly, and returning to combinatorics, it is interesting to appreciate the difficulty of finding an interpretation for a combinatorial deduction in the case of the Rayleigh-Wien law \eqref{comx}. $W$ turns into:
$$W=\frac{p!}{(p-n)!n!},$$
which in that limit coincides with distributing $n$ modes in $p$ quanta:
$$W=\frac{p^{n}}{n!}.$$
Nothing too convincing. Combinatorial analysis only fits when radiation shows corpuscular behavior. So far, these are the first appearances of corpuscularity caused by the use of combinatorics. It will be more than ten years before both quantum statistics and duality emerge. However, we already saw that in his crucial contribution of 1909, Einstein obtained an expression for the mean value of the fluctuations of the radiation energy \eqref{fluctu}. In that expression, the first term corresponds to the corpuscular nature of light, while the second one corresponds to the ondulatory nature of light. Hence, he concluded ``the current theory of radiation is incompatible with this result'' (\cite{beck2}. p. 365). This is an interesting (and original) way of showing again the kind of behaviors of radiation. Note, furthermore, that it was not until a few years later that Lorentz demonstrated that the second term did indeed correspond to the fluctuations of a cavity with classical electromagnetic waves (\cite{duncanjanssen2019}, p. 123).

Finally, although it falls slighty outside our tine frame, we want to mention the contribution of 1916 by which Einstein proposed the so-called Einstein coefficients \citep{einstein1916}. With this contribution he established a link between Planck's law and the Bohr atom which the Danish physicist later incorporated into his theory (\cite{kragh}, p. 199). Considering two states of a gas molecule $Z_n$ and $Z_m$ ($\varepsilon_m > \varepsilon_n$), the molecule can go from state $Z_n$ to state $Z_m$ absorbing a quanta of energy $\varepsilon_m - \varepsilon_n$ and from state $Z_m$ to $Z_n$ by emitting energy of the same value. Einstein introduced three kinds of processes for determining the statistical law followed by the transition between these two states: the spontaneous emission and the induced emission or absorption.

On the one hand, Einstein defines the probability that a molecule transitions from state $Z_m$ to a lower energy state $Z_n$ by the spontaneous emission of radiation of energy $\varepsilon_m -\varepsilon_n $ occurs as: 

\begin{equation}
    dW = A_{m}^n dt
\end{equation}
where $A_{m}^{n}$ is a constant that depends on the two states of the transition. On the other hand, the molecule can change from state $Z_n$ to the state $Z_m$ by absorbing energy and viceversa, and those probabilities would be:\footnote{As is known, this is the theoretical basis of the operation of the laser, which would still take a few years to be designed.}

\begin{equation}
    dW = B_{n}^m \rho dt
\end{equation}

\begin{equation}
    dW = B_{m}^n \rho dt.
\end{equation}
Einstein considers that for the radiation density being constant (in equilibrium), there should be as many processes of emission as of absorption. That is:

\begin{equation} \label{trans}
    p_n e^{- \beta \varepsilon_n} B_n^m \rho = p_m e^{-\beta \varepsilon_m} (B_{m}^n \rho + A_m^n).
\end{equation}
Note that he uses Boltzmann factor for the distribution of energies. Furthermore, since $p_n B_n^m$ must equal $p_m B_m^n$ in the whole range, the dynamic equilibrium condition is:

\begin{equation} \label{ppla}
    \rho = \frac{ \frac{A_m^n}{B_m^n} }{ e^{-\beta (\varepsilon_m - \varepsilon_n)} - 1 },
\end{equation}
which is the behavior of radiation density in Planck's law if we identify:

$$\frac{A_m^n}{B_m^n}=\frac{8 \pi h\nu^{3}}{c^{3}}.$$

To study the low-density radiation limit (i.e. Wien's limit), we simply turn to equation \eqref{trans} and see that we obtain the same result as if we applied the limit directly to formula \eqref{ppla}. In that range, spontaneous emission would dominate over induced emission. Now, what happens at the other extreme? In the expression \eqref{trans} we should neglect the spontaneous emission. However, we do not get anywhere. Classically, then, quantum approach through Bohr-type transitions does not seem to make sense. Corpuscularity of light emerges as linked to quantization of atomic energy levels.

\subsection{Photons, the modern light quanta}

As it is widely known, the reluctance regarding the existence of light quanta was initially justified even after the main experimental characteristics of the photoelectric effect were measured and demonstrated to be in accordance with the predictions of Einstein's quantum theory \citep{stuewerphot}. Robert Millikan worked for years in his laboratory at the University of Chicago on the photoelectric effect, using light in the visible spectrum—the spectrum of mercury—and various alkali metals as targets.\footnote{On the early history of photoelectric effect see, for instance, \cite{duncanjanssen2019}, p. 103.}  He presented his results in a paper of 1916 detailing his experiments and the results obtained, confirming Einstein's law ($E_{m}$ is the maximum kinetic energy of the electron and $W_{\circ}$ the work function of the metallic cell): 

$$E_m = h \nu - W_{\circ}.$$
Millikan, however, did not believe that his results confirmed the existence of light quanta (\cite{millikan}, p. 230). Millikan's position was not unusual. After all, Einstein's intention was not so much to establish a new theory as to question the current one. No significant findings were made until Arthur Compton and Peter Debye derived the relativistic kinematics for the scattering of a photon from a rest electron.\footnote{On the history and relevance of Compton effect, see \cite{stuewer}.} 

In the same paper where Compton develops the conservation laws for this kind of collisions, we find the experimental results that led him to Einstein's idea, but, contrary to Debye, without mentioning him (\cite{compton}, p. 501):

\begin{quote}
The experimental support for the theory indicates very convincingly that a quantum of radiation carries with it momentum as well as energy.
\end{quote}

Einstein himself, with theoretical arguments, had discovered the momentum of the photon in 1916, through a masterful new use of fluctuations in a \textit{Gedankenexperiment } \citep{einstein1916}. It was in the same paper that he discussed the transition amplitudes mentioned above.

The typical ranges of these experiments are the visible, in the case of the photoelectric effect, and X-rays, in the Compton case. Hence, the tools of classical physics led, in one way or another, to the conclusion that electromagnetic radiation exhibited corpuscular behavior far in the visible range and beyond.

Currently, photons are regarded, according to QFT, as the quantization of the excitations of the electromagnetic field. Photons are massless bosons with spin $1$ and responsible of all electric and magnetic fields. The behaviour of the electromagnetic field is described by Maxwell equations, which are the massless case of the more general vector particle equations: the Proca-de Broglie equation. 

Semiclassical theories where light was not quantized were mainly developed until 1970's, when the photon antibunching proved to be a crucial experiment for the corpuscular nature of light \citep{loudon} Although nowadays there are plenty experiments that show the corpuscular behavior of light --from vacuum fluctuations to many particle correlations \citep{muthukrishan,hentschel,shalm}-- we think antibunching illustrates well the quantization in the emission process. Recall that Einstein hypothesis was originally meant to tackle with emission and absorption processes \citep{einstein1905}. In particular, the antibunching was deeply studied in 1977 by H. Jeff Kimble, Mario Dagenais and Leonard Mandel \citep{kimble} using resonance fluorescence from sodium atoms excited ($\lambda \approx 589$ nm). The significance of this observation is precisely described by them (\cite{kimble}, p. 691):

\begin{quote}
It is pointed out that, unlike photoelectric bunching, which can be given a semiclassical interpretation, antibunching is understandable only in terms of a quantized electromagnetic field. The measurement also provides rather direct evidence for an atom undergoing a quantum jump.
\end{quote}

In a nutshell, photon antibunching is simply a light field where photons are more spaced than in a coherent wave, i.e. than in a classical light wave. Therefore, in an antibunching experiment, photons are detected with a lower probability than in a classical light wave. Moreover, ``the quantum nature of the radiation field and the quantum jump in emission, which are inextricably connected, are therefore both manifest'' (\cite{kimble}, p. 694) in the measurement.

In order to understand the antibunching phenomena, it is interesting to see that the second order correlation function $g^{(2)} (\tau)$ describes in this case the probability of detecting two photons in a $\tau$ time. In the case of a light beam whose photons are detected in bunches, the probability density of detecting photons at two times $t$ and $t+\tau$ is greater when $\tau \rightarrow 0$ and is lower when $\tau$ is greater than the coherence time. In particular, if $g^{(2)} (\tau) = 1$, we have that photons are not correlated; while for $g^{(2)} (\tau) > 1$ photons arrive in bunches. This last case is due to the nature of the emissor.

The second order correlation function can be computed as: 

\begin{equation}
g^{(2)} (\tau) = \frac{ \langle I(t) I(t+\tau) \rangle  }{\langle I(t)\rangle^{2} }
\end{equation}
where $I (t)$ is the intensity at time $t$, $\tau$ is the time between measurements and the terms in brackets $\langle ... \rangle$ refers to the average of the quantities. Therefore, it is simple to understand that if no photon is detected at one of both times $t$ and $t + \tau$, the value of the numerator will be zero. Nonetheless, generally in the antibunching cases it will yield lower value than one. 

An ideal unique quantum source would have $g^{(2)} (\tau = 0)=0$, although $g^{(2)} (\tau) < 1$ is valid for proving a quantum behaviour of a source, not necessarily ideal nor unique. Although the original experiment that proved the existence of quanta corresponded to the wavelength of the transition of sodium atoms \citep{kimble}, the antibunching has been proven for quantum emitters across the electromagnetic spectrum. On the one hand, for high frequency, the photon antibunching was proven by \citep{diedrich1987}, who used a Magnesium ion for the ultraviolet region. On the other hand, \citep{michler2000} proved the photon antibunching in the infrared region.

From the QFT point of view, the world is a quantized field and it can be studied classically just in some limits. A classical field is simply a regular function that can be written through a Fourier transform to the wave vector/frequency space, i.e. a linear combination of modes with those two magnitudes well-defined everywhere (\cite{duncan2012}, p. 223). The quantization implies a complementarity between them. There are divergences in the integrals when computing a Feynman diagram due to having a very large energy (ultraviolet divergence) or a very low energy (infrared divergence).  This second case does only happen for massless particles --such as photons-- and soft photons are virtual particles employed for being able to solve the integrals without divergences.

A quantum field can be considered classical whenever it has a high occupation number for the individual modes, something that can only happen in bosonic fields --since for fermions the Pauli exclusion principle applies-- associated to stable particles --that ensure the classical configuration for a classical time-scale-- and a non-interacting limit (\cite{duncan2012}, p. 228). There are only two stable boson particles that satisfy these requirements: the photon and the gluon, but the color confinement implies a relation between the physical state and the local fields (quarks and gluons) in such a way that gluons do not fulfill the non-interacting limit condition. Therefore, the electromagnetic field is the only one with a classical limit. 

That is, according to QFT, the relevant parameter for characterizing a classical behavior is the occupation number. High occupation number will yield to the classical limit --the ondulatory behaviour. Note that the low occupation number is the low radiation intensity Einstein referred to in 1905 in order to justify the introduction of light quanta. In any case, nowadays the photon is defined equally across the entire electromagnetic spectrum. Its origin in the highest energy range is related only to its statistical properties, as that was the context in which Einstein conceived of it.

\section{Final remarks}

Einstein’s heuristic argument of 1905 was met with skepticism by his contemporaries, who largely ignored it—a reaction later shared by historians of physics.  

This distrust is understandable. On one hand, the wave properties of radiation (particularly interference) were well-established, and reintroducing a corpuscular view to explain phenomena like the photoelectric effect seemed unwarranted. Importantly, Einstein did not aim to advocate for a purely particle-like conception of light but rather to question whether the wave theory alone could fully account for all observed phenomena. For him, Maxwell’s theory could not be the final word.  

Another reason for skepticism was the limited understanding and acceptance of statistical methods at the time. It was only during the first quarter of the 20th century that the statistical approaches of Boltzmann and Gibbs became integral to physics. Einstein himself, at the Solvay Conference, suggested that quantum theory might help clarify Boltzmann’s principle—a principle whose very meaning and utility were still under debate. Issues such as the factor $N!$ and the entropy constant also contributed to these discussions, though we have not explored them here. As already mentioned, in those years there was nothing like an established and accepted ``classical'' doctrine (\cite{staley}, p. 347). In a future work we will extend our analysis focused on the Boltzmann principle to the more general panorama of statistical mechanics in the first third of the 20-th century.

Even within the framework of statistical mechanics, Einstein’s argument was unconventional. Applied to gases, it reverses the usual explanation of irreversibility. Consider a gas initially confined to a volume $v_{\circ}$ that freely expands into a larger volume $v$: the probability of it spontaneously returning to $v_{\circ}$ is negligible. While the simplicity of an ideal gas allows a comparison of probabilities, extending this reasoning to radiation is far less straightforward. What does it mean for monochromatic radiation to concentrate instantaneously in a portion of the volume? Such a process is not just unimaginable—its probability is incalculable. This is why Einstein inverted Boltzmann’s principle, deriving probability from entropy. But this approach required the system to remain in equilibrium at all times.  

Ehrenfest reinterpreted Einstein’s argument as describing free expansion instead, allowing a comparison of equilibrium-state entropies. In his analysis of cavity radiation, Ehrenfest used adiabatic transformations to track how frequency (and energy) varied over time in continuous transformations, not in fluctuations.  

Crucially, entropy comparisons at a certain temperature $T$ are only meaningful for equilibrium states. For an ideal gas, this poses no issue, since its energy is volume-independent and proportional to $T$. Some critics have dismissed Einstein’s argument as tautological, claiming that equating radiation’s entropy with that of an ideal gas implicitly assumes a corpuscular nature. However, this critique is flawed. While the volume dependence is indeed identical, the analogy breaks down because the two states of radiation being compared do not share the same temperature—a consequence of the absence of a fixed particle number constraint.  

Our goal is not to identify errors in Einstein’s reasoning, as he himself never presented it as a rigorous proof. Rather, we argue that the evolution of his ideas reflects the unsettled state of statistical physics in the early days of quantum theory. Statistical methods were neither well understood nor firmly established. By 1924, when Bose derived Planck’s law, Einstein had to revisit the use of complexions—a concept he had previously avoided. The absence of a ``molecular-theoretical'' model made it impossible to compute radiation’s entropy as Boltzmann had done for gases. Planck attempted this, but Einstein criticized his approach for its fatal ambiguity in defining probability.  

Bose’s serendipitous derivation of the quantum ideal gas law—later refined by Einstein—highlighted the need for a mechanical model underlying Bose’s complexions. The interdependence of molecules led Einstein to consider De Broglie’s hypothesis. Yet, his wariness of combinatorics drove him to explore gas properties without relying on them—as seen in his lesser-known third paper on quantum gases, which sought arguments independent of ``incriminated'' statistics \citep{perezsauer}.  

The episode we have discussed is a good example of the exploratory methodology followed by Einstein. We have detected what we could call up to three methodological inversions of different kinds:

\begin{enumerate}
\item In Boltzmann's principle, obtaining $W$ from $S$ and not the other way around (1905, 1909).
\item Basing Boltzmann's principle on the quantum hypothesis and not what was usual even in his own works: starting from the Boltzmann principle to explore the quantum hypothesis (1911, 1914).
\item And in the 1920's, deducing a molecular model from $W$ and not try to calculate $W$ from a certain model (1925).
\end{enumerate}

Beyond this discussion, we have also shown how historical analysis can deepen our understanding of physical concepts. Einstein’s (and Ehrenfest’s) work between 1905 and 1915 revealed how, according to statistics, radiation behaves differently across the spectrum: particle-like properties emerge at low densities. Does this imply, for instance, that photons beyond the infrared are meaningless? QFT asserts that photons —excitations of the electromagnetic field, not classical particles— exist across the entire spectrum. The \textit{low density} condition Einstein referenced in 1905 corresponds to \textit{low occupancy} in QFT. Classical approximations only hold under high occupancy. Modern antibunching experiments—where radiation is so sparse that semiclassical models fail—provide the most compelling evidence for Einstein’s 1905 heuristic hypothesis: Maxwell’s theory, though successful, required completion. This completion ultimately transformed our understanding of radiation. While the Wien limit once exhibited anomalous behavior, today it is the Rayleigh-Jeans limit—justified by high occupancy—that shows peculiarity.  

\vspace{3ex}
\small
\flushleft
\textbf{Acknowledgements}
Part of this research has been funded by the Spanish \textit{Ministerio de Ciencia, Innovaci\'on y Universidades} under contract PID2023-147710NB-I00. We would like to especially thank Tony Marzoa for his comments (the result of a careful and critical reading of a first version of the manuscript) and the amount of bibliography provided. 


\bibliographystyle{elsarticle-harv}

\begin{thebibliography}{99}


\bibitem[Beck(1990)]{beck2}
    Beck, Anne (1990). \textit{The Collected Papers of Albert Einstein. Volume 2: The Swiss Years: Writings, 1902-1909. English translation supplement}. Princeton University Press.  

\bibitem[Beck(1994)]{beck3}
    Beck, Anne (1994). \textit{The Collected Papers of Albert Einstein. Volume 3: The Swiss Years: Writings, 1909-1911. English translation supplement}. Princeton University Press.  

\bibitem[Beck(1995)]{beck5}
    Beck, Anne (1995). \textit{The Collected Papers of Albert Einstein. Volume 5: The Swiss Years: Correspondence, 1902-1914. English translation supplement}. Princeton University Press. 
        
\bibitem[Compton(1923)]{compton}
    Compton, Arthur (1923). ``A Quantum Theory of the Scattering of X-rays by Light Elements.'' Physical Review 21, 483-502.

\bibitem[Darrigol(1991)]{darrigol}
    Darrigol. Olivier (1991). ``Statistics and combinatorics in early quantum theory, II: Early symptoma of indistinguishability and holism''. Historical Studies in the Physical and Biological Sciences 21, 237-298.

\bibitem[Desalvo(1992)]{desalvo}
    Desalvo. Agostino (1992). ``From the chemical constant to quantum statistics: A thermodynamic route to quantum mechanics.'' Physis 29, 465-537.

\bibitem[Dorling(1971)]{dorling}
    Dorling, Jon (1971). ``Einstein's Introduction of Photons: Arguments by Analogy or Deduction from the Phenomena?'' British Journal for the Philosophy of Science 22, 1-8.

\bibitem[Diedrich \& Walther(1987)]{diedrich1987}
  Diedrich, Frank and Walther, Herbert (1987). ``Nonclassical radiation of a single stored ion'' Physical Review Letters 58, 203-206.


\bibitem[Duncan(2012)]{duncan2012}
    Duncan, Anthony (2012). \textit{The Conceptual Framework of Quantum Field Theory}. Oxford University Press.

\bibitem[Duncan \& Janssen(2019a)]{duncanjanssen2019}
    Duncan, Anthony and Janssen, Michel (2019a). \textit{Constructing Quantum Mechanics. Volume 1: The Scaffold, 1900-1923.} Oxford University Press.

\bibitem[Duncan \& Janssen(2019b)]{duncanjanssen2019w}
    Duncan, Anthony and Janssen, Michel (2019b). ``The (first) Einstein fluctuation theorem. Web Resource''. Web Resource of \citep{duncanjanssen2019}

\bibitem[Ehrenfest(1911)]{ehrenfest1911}
    Ehrenfest, Paul (1911): ``Welche Züge der Lichtquantenhypothese spielen in der Theorie der Wärmestrahlung eine wesentliche Rolle?''. Annalen der Physik 36, 91–118. Reprinted in \cite{klein}, pp. 185-212. 

\bibitem[Ehrenfest \& Kamerlingh-Onnes(1911)]{ehrkam}
    Ehrenfest, Paul and Kamerlingh-Onnes, Heike (1914): ``Simplified deduction of the formula from the theory of combinations which Planck uses as the basis of his radiation theory''. Proceedings of the Amsterdam Academy 17, 870-873. Reprinted in \cite{klein}, pp. 353-356.

\bibitem[Einstein(1902)]{einstein1902}
    Einstein, Albert (1902). ``Kinetische Theorie des Wärmegleichgewicht und des zweiten Hauptsatzes der Thermodynamik.'' Annalen der Physik 9, 417-433. Reprinted in \cite{stachel1989}, pp. 56-75. English version in \cite{beck2}, pp. 30-47.

\bibitem[Einstein(1903)]{einstein1903}
    Einstein, Albert (1903). ``Eine Theorie der Grundlagen der Thermodynamik''. Annalen der Physik 11, 170-187. Reprinted in \cite{stachel1989}, pp. 76-97. English version in \cite{beck2}, pp. 48-67.
    
\bibitem[Einstein(1904)]{einstein1904}
    Einstein, Albert (1904). ``Zur allgemeinen molekularen Theorie der Wärme.''. Annalen der Physik 14, 354-362. Reprinted in \cite{stachel1989}, pp. 98-108. English version in \cite{beck2}, pp. 68-77.
    
\bibitem[Einstein(1905)]{einstein1905}
    Einstein, Albert (1905). ``Über einen die Erzeugung und Verwandlung des Lichtes betreffenden heuristischen Gesichtpunkt''. Annalen der Physik 17, 132-148. Reprinted in \cite{stachel1989}, pp. 149-169. English version in \cite{beck2}, pp. 86-103.

\bibitem[Einstein(1907a)]{einstein1907}
    Einstein, Albert (1907a). ``Über die Gültigkeitsgrenze des Satzes von thermodynamischen Gleichgewicht und über die Möglichkeit einer neuen Bestimmung der Elementarquanta''. Annalen der Physik 22, 569-572. Reprinted in \cite{stachel1989}, pp. 392-397. English version in \cite{beck2}, pp. 225-228.

\bibitem[Einstein(1907b)]{einstein1907sh}
    Einstein, Albert (1907b). ``Die Plancksche Theorie der Strahlung un die Theorie der spezifischen Wärme''. Annalen der Physik, 22, 180–190. Reprinted in \cite{stachel1989}, pp. 378-391. English version in \cite{beck2}, pp. 214–224.
    
\bibitem[Einstein(1909a)]{einstein1909}
    Einstein, Albert (1909a). ``Zum gegenwärtigen Stand des Strahlungsproblems''. Physikalische Zeitschrift 10, 185-193. Reprinted in \cite{stachel1989}, pp. 541-555. English version in \cite{beck2}, pp. 357-375.

\bibitem[Einstein(1909b)]{einstein1909b}
    Einstein, Albert (1909b). ``Über die Eintwickelung unserer Anschauungen über das Wesen und die Konstitution der Strahlung''. Deutsche Physikalische Gesellschaft, Verhandlungen 7, 482-500. Reprinted in \cite{stachel1989}, pp. 563-583. English version in \citep{beck2}, pp. 379-394.

\bibitem[Einstein(1910a)]{einstein1910}
    Einstein, Albert (1910a). ``Theorie der Opaleszenz von homogenen Flüssigkeiten und Flüssigkeitsgemischen in der Nähe des kritischen Zustand''. Annalen der Physik 33, 1275-1298. Reprinted in \cite{kleinkox}, pp. 286-312. English version in \cite{beck3}, pp. 231-249.

\bibitem[Einstein(1910b)]{einstein1910b}
    Einstein, Albert (1910b). ``Sur la théorie des quantités lumineuses et la question de la localisation de l'énergie électromagnétique'' Archives des sciences physiques et naturelles 29, 525-528. Reprinted in \cite{kleinkox}, pp. 248-253. English version in \cite{beck3}, pp. 207-208.

\bibitem[Einstein(1914)]{einstein1914}
    Einstein, Albert (1914). ``Beiträge zeu Quantentheorie''. Deutsche Physikalische Gesellschaft. Verhandlungen 16: 820-828. Reprinted in \cite{kleinkox2}, pp. 29-40. English version in \cite{engel1997}, pp. 20-26.     

\bibitem[Einstein(1917)]{einstein1916}
    Einstein, Albert (1916). ``Zur Quantentheorie der Strahlung''. Physikalische Zeitschrift, 18 (1917), 121-128. Reprinted in \cite{kleinkox2}, pp. 381-398. English version in \cite{engel1997}, pp. 220-233. 

        
\bibitem[Engel(1997)]{engel1997}
    Engel, Alfred (1997). \textit{The collected papers of Albert Einstein, vol. 6. The Berlin years: Writings, 1914-1917. English translation of selected texts.}Princeton University Press.

\bibitem[Gibbs(1902)]{gibbs}
    Gibbs, Josiah Willard (1900). \textit{Elementary Principles in Statistical Mechanics. }Yale University Press.

\bibitem[Hentschel(1998)]{hentscheleins}
    Hentschel, Ann M. (1999). \textit{The Collected Papers of Albert Einstein. Volume 8. Part A: The Berlin Years: Correspondence, 1914-1917. English translation supplement}. Princeton University Press. 
       
\bibitem[Hentschel(2018)]{hentschel}
    Hentschel, Klaus (2018). \textit{The History and Mental Models of Light Quanta.} Springer.


\bibitem[Irons(2004)]{irons}
    Irons, F. E. (2004). ``Reappraising Einstein's 1905 application of thermodynamics and statistics to radiation''. European Journal of Physics 25, 269-277.

\bibitem[Kimble et al.(1977)]{kimble}
    Kimble, H. Jeff, Dagenais, Mario and Mandel, Leonard (1977). ``Photon Antibunching in Resonance Fluorescence''. Physical Review Letters 39, 691-695.

\bibitem[Klein(1959)]{klein}
    Klein, Martin J. (ed.) (1959). \textit{Paul Ehrenfest. Collected scientific papers (with an introduction by H.B.G. Casimir)}. North–Holland.

\bibitem[Klein et al.(1993)]{kleinkox}
    Klein, Martin J.; Kox, Anne J.; and Schulmann, Robert. (1989). \textit{The Collected Papers of Albert Einstein. Volume 3: The Swiss Years: Writings, 1909-1911).} Princeton University Press. English translation in \cite{beck3}.   

\bibitem[Klein et al.(1993)]{kleinkoxlet}
    Klein, Martin J., Kox, Anne J. and and Schulmann, Robert. (1989). \textit{The Collected Papers of Albert Einstein. Volume 5: The Swiss Years: Correspondence, 1902-1914).} Princeton University Press. English translation in \cite{beck5}.   

\bibitem[Klein et al.(1996)]{kleinkox2}
    Klein, Martin J., Kox, Anne J. and and Schulmann, Robert. (1989). \textit{The Collected Papers of Albert Einstein. Volume 6: The Berlin Years: Writings, 1914-1917).} Princeton University Press. English translation in \cite{engel1997}. 
   
\bibitem[Kojevnikov(2002)]{kojevnikov02}
  Kojevnikov, Alexei (2002). ``Einstein's Fluctuation Formula and the Wave-Particle Duality''. In Balashov, Y. and Vizgin, V. (eds.) Einstein Studies in Rusia. Einstein Studies, vol. 10. Brikhäuser, pp. 181-228.

\bibitem[Kragh(2012)]{kragh}
    Kragh, Helge (2012). \textit{Niels Bohr and the Quantum Atom. The Bohr Model of Atomic Structure 1913-1925.} Oxford University Press.
    
\bibitem[Kuhn(1978)]{kuhn}
    Kuhn, Thomas S. (1978). \textit{Black-body theory and the quantum discontinuity, 1894-1912.} Clarendon Press.

\bibitem[Lamb(1995)]{lamb}
    Lamb, Willis E. Jr. (1995). ``Anti-photon''. Applied Physics B 60, 77-84.

\bibitem[Langevin \& De Broglie(1912)]{solvay}
    Langevin, Paul and De Broglie, Maurice (1912) (eds.). \textit{La theorie du rayonnement et les quanta.} Gauthier-Villars.

\bibitem[Loudon(1976)]{loudon}
    Loudon, R. (1976). ``Photon Bunching and Antibunching''. Physics Bulletin 27, 21-23.
    
\bibitem[Michler et al.(2000)]{michler2000}
  Michler, Peter; Kiraz, Alper; Becher, Cristoph; Schoenfeld, Winston V.; Petroff,  Pierre M.; Lidong Zhang, E. Hu and Imamoglu, Atac (2000). ``A Quantum Dot Single-Photon Turnstile Device''. Science 290, 2282-2285.

\bibitem[Millikan(1917)]{millikan}
    Millikan, Robert (1917). \textit{The Electron: Its Isolation and Measurement and the Determination of Some of its Properties}. University of Chicago Press.

\bibitem[Monaldi(2009)]{monaldi}
    Monaldi, Daniela (2009). ``A note on the prehistory of indistinguishabilty particles''. Studies in History and Philosophy of Modern Physics 40, 383-394.
    
\bibitem[Muthukrischan et al.(2003)]{muthukrishan}
    Muthukrishnan, Ashok, Scully, Marlan O. and M. Zubairy, Suhail (2003). ``The Concept of the Photon—Revisited''. Optics \& Photonics News 14, 18-27.    
    
\bibitem[Navarro(2023)]{navarro}
    Navarro, Luis (2023). \textit{The Lesser-Known Albert Einstein}. Springer.

\bibitem[Navarro \& Pérez(2002)]{navarroperez2002}
    Navarro, Luis and Pérez, Enric (2002). ``Principio de Boltzmann y primeras ideas cuánticas en Einstein''. Dynamis 22, 377-410.

\bibitem[Navarro \& Pérez(2004)]{navarroperez2004}          
    Navarro, Luis and Pérez, Enric (2004). ``Paul Ehrenfest on the necessity of quanta (1911): Discontinuity, quantization, corpuscularity, and adiabatic invariance''. Archive for History of Exact Sciences 58, 97-141. 

\bibitem[Navarro \& Pérez(2006)]{navarroperez2006}          
    Navarro, Luis and Pérez, Enric (2006). ``Paul Ehrenfest: The genesis of the adiabatic hypothesis, 1911-1914''. Archive for History of Exact Sciences 60, 209-267.

\bibitem[Norton(2006)]{norton}
    Norton, John (2006). ``Atoms, entropy, quanta: Einstein's miraculous argument of 1905''. Studies in History and Philosophy of Modern Physics 37, 71-100.

\bibitem[Pais(1982)]{pais}
    Pais, Abraham (1982). \textit{`Subtle is the Lord...' The Science and the life of Albert Einstein.} Oxford University Press.

\bibitem[Pérez \& Sauer(2010)]{perezsauer}
    Pérez, Enric and Sauer, Tilman (2010). ``Einstein's quantum theory of the monatomic ideal gas: non-statistical arguments for a new statistics.'' Archive for History of Exact Sciences 64, 561-612.

\bibitem[Pérez \& Ibáñez(2022)]{pereziban}
    Pérez, Enric and Ibáñez, Joana (2022). ``Indistinguishable elements in the origins of quantum statistics. The case of fermi-Dirac statistics. European Physical Journal H 47, 1.

\bibitem[Planck(1991)]{planck}
    Planck, Max (1991). \textit{The Theory of Heat Radiation}. Dover. Unabridged republication of the Dover edition published in 1959. Translation by M. Masius of the second edition of \textit{Wärmestrahlung}, originally published in 1914.
    
\bibitem[Provost \& Bracco(2008)]{provostbracco}
    Provost, Jean-Pierre and Bracco, Christian (2008). ``Einstein's quanta and the 'true' volume dependence of the black-body entropy''. European Journal of Physics 29, 1085-1090.


\bibitem[Rashkovskiy(2016)]{rashkovskiy}
    Rashkovskiy, Sergey A. (2016). ``Quantum mechanics without quanta: the nature of the wave–particle duality of light''. Quantum Studies: Mathematics and Foundations 3, 147–160.
    
\bibitem[Rieckers(2025)]{rieckers}
    Rieckers Alfred (2025). ``An Actual Discussion of Einstein’s Early Fusion Picture for Photons and Classical Maxwell Fields''. International Journal of Theoretical Physics 64, 61. 

\bibitem[Rynasiewicz \& Renn(2006)]{rynorenn}
    Rynasiewicz, Robert and Renn, Jürgen (2006). ``The turning point for Einstein's \textit{Annus mirabilis}''. Studies in History and Philosophy of Modern Physics 37, 5-35.
    
\bibitem[Shalm et al.(1998)]{shalm}
    Shalm, Lynden K.; Steinberg, Aephraim M.; Kwiat, Paul G. and Chiao, Raymond Y. (2023). Quantum Optical Tests of the Foundations of Physics. In: Drake, G.W.F. (eds) \textit{Springer Handbook of Atomic, Molecular, and Optical Physics}. Springer Handbooks. Springer.

\bibitem[Schulmann et al.(1998)]{schulmannl}
    Schulmann, Robert, Janssen, Michel and Illy, József (1998). \textit{The Collected Papers of Albert Einstein. Volume 8. Part A: The Berlin Years: Correspondence, 1914-1917).} Princeton University Press. English translation in \cite{hentscheleins}.   

\bibitem[Smith \& Seth(2020)]{smithseth}
    Smith, George and Seth, Raghvav (2020). \textit{Brownian Motion and Molecular Reality. A Study in Theory-Mediated Measurement.} Oxford University Press.
    
\bibitem[Speziali(1972)]{speziali}
    Speziali, Pierre (1972). \textit{Albert Einstein, Michele Besso. Correspondance.}  Hermann.
    
\bibitem[Stachel et al.(1989)]{stachel1989}
    Stachel, John, Cassidy, David C., Renn, Jürgen and Schulmann, Robert. (1989). \textit{The Collected Papers of Albert Einstein. Volume 2: The Swiss Years: Writings, 1902-1909).} Princeton, Princeton University Press. English translation in \cite{beck2}.

\bibitem[Staley(2008)]{staley}
    Staley, Richard (2008). \textit{Einstein's generation. The origins of the Relativity Revolution.} Chicago University Press.

\bibitem[Stuewer(1975)]{stuewer}
    Stuewer, Roger (1975). \textit{The Compton effect: turning point in physics.} New York, Science History Publications.

\bibitem[Stuewer(2006)]{stuewerphot}
    Stuewer, Roger (2006). ``Einstein's revolutionary light-quantum hypothesis.'' Acta Physica Polonica B 37, 543-558.

\bibitem[Taschetto(2025)]{taschetto}
    Taschetto, Diana (2025). ``Rewriting the Quantum ``Revolution'' ''. Studies in History and Philosophy of Science 109, 72-88. 
    
\bibitem[Uffink(2006)]{uffink}
    Uffink, Jos (2006). ``Insuperable difficulties: Einstein's statistical road to molecular physics''. Studies in History and Philosophy of Modern Physics 37, 36-70.

\bibitem[Villas-Boas et al.(2025)]{villasboas}
   Villas-Boas, Celso J.; M\'aximo, Carlos E.; Paulino, Paulo J.; Bachelard, Romain P. and Rempe, Gerhard (2025). ``Bright and Dark States of Light: The Quantum Origin of Classical Interference''. Physical Review Letters 134:  13.

\end{thebibliography}

\end{document}